\documentclass[10pt]{article}

\usepackage[margin=1in]{geometry}
\usepackage{setspace}
\usepackage{times} 
\usepackage{cite}
\usepackage{amsmath,amssymb,amsfonts}
\usepackage{algorithm}
\usepackage{algorithmic}
\usepackage{graphicx}
\usepackage{xr}
\usepackage{hyperref}
\usepackage{textcomp}
\usepackage{placeins}
\usepackage{xcolor}
\usepackage{ifthen}

\newboolean{showhighlight}
\setboolean{showhighlight}{false} 
\newcommand{\hl}[1]{\ifthenelse{\boolean{showhighlight}}{\textcolor{blue}{#1}}{#1}}

\title{\textbf{Accelerating Low-field MRI: From Compressed Sensing to Deep Learning Reconstruction with CNNs and Transformers}}

\author{Efrat Shimron,$^{1,2}$ Shanshan Shan,$^{3,4}$ James Grover,$^{4}$, Neha Koonjoo,$^{5,6}$\\ Sheng Shen,$^{5,6}$ Thomas Boele,$^{4,5}$ Annabel J. Sorby-Adams,$^{6,7}$\\ John E. Kirsch,$^{5,6}$ Matthew S. Rosen,$^{6,7,8}$ and David E. J. Waddington$^{4\ast}$\\
	\\
    \parbox{1.0\linewidth}{\centering\normalsize{$^{1}$} Technion - Israel Institute of Technology, Department of Electrical and Computer Engineering and Department of Biomedical Engineering, Haifa, Israel}\\
    \parbox{1.0\linewidth}{\centering\normalsize{$^{2}$} Technion -  Israel Institute of Technology, May-Blum-Dahl Human MRI Research Center, Haifa, Israel}\\
    \parbox{1.0\linewidth}{\centering\normalsize{$^{3}$}  Center for Molecular Imaging and Nuclear Medicine, State Key Laboratory of Radiation Medicine and Protection, School for Radiological and Interdisciplinary Sciences (RAD-X), Soochow University, Suzhou 215123, China}\\
	\parbox{1.0\linewidth}{\centering\normalsize{$^{4}$Image X Institute, Sydney School of Health Sciences, Faculty of Medicine and Health, University of Sydney, NSW 2006, Australia}}\\        
    \parbox{1.0\linewidth}{\centering\normalsize{$^{5}$A. A. Martinos Center for Biomedical Imaging, 149 Thirteenth St., Charlestown, MA 02129, USA}}\\
	\parbox{1.0\linewidth}{\centering\normalsize{$^{6}$Harvard Medical School, 25 Shattuck St., Boston, MA 02115, USA}}\\
	\parbox{1.0\linewidth}{\centering\normalsize{$^{7}$Department of Neurology and the Center for Genomic Medicine, Massachusetts General Hospital, Boston, MA 02114, USA}}\\
    \parbox{1.0\linewidth}{\centering\normalsize{$^{8}$Department of Physics, Harvard University, 17 Oxford St., Cambridge, MA 02138, USA}}\\
	\normalsize{$^\ast$To whom correspondence should be addressed; E-mail: david.waddington@sydney.edu.au}
}

\date{} 

\begin{document}

\onehalfspacing
\maketitle

\begin{abstract}
Portable, low-field Magnetic Resonance Imaging (MRI) scanners are increasingly being deployed in clinical settings. However, key barriers to their widespread use include low signal-to-noise ratio (SNR), generally low image quality, and long scan durations. Hence, methods for accelerating acquisition and boosting image quality are critically important to enable clinically actionable, high-quality imaging in these systems. Despite the role that compressed sensing (CS) and \hl{deep learning (DL)}-based methods have played in improving image quality for high-field MRI, their adoption for low-field imaging is still in its infancy, and it remains unclear how robust these methods are in low-SNR regimes.

Here, we propose, investigate, and compare \hl{four} reconstruction approaches: (i) L1-wavelet CS; (ii) a data-driven network; (iii) an unrolled network; and \hl{(iv) a Swin Transformer Cascade}. We evaluate their performance across a range of SNR values using publicly available datasets and ultra-low field (6.5~mT) MRI data. \hl{Our results show that the unrolled network and Swin Transformer cascade outperform CS and data-driven models. While transformer-based models achieve the highest performance at high SNR, unrolled convolution-based networks are more robust in ultra-low SNR settings and often outperform transformers—indicating that simpler DL architectures may be better suited to low-field MRI.}

This work highlights both the potential and limitations of advanced reconstruction techniques in low-field MRI and pinpoints effective DL strategies for addressing SNR challenges.
\end{abstract}

\section{Introduction}
\label{sec:introduction}
Low-field Magnetic Resonance Imaging (MRI) is revolutionizing medical imaging by offering cost-effective and portable solutions, yet it faces significant technical challenges. While high-field MRI systems (1.5-3 Tesla) are widely used for their high image quality, low-field MRI systems (0.005-0.1 Tesla) are emerging as a more accessible alternative, particularly in resource-limited settings \cite{Sarracanie2015,Kimberly2023,Zhao2024,Salameh2023}. Low-field MRI has already shown promise in diagnosing brain diseases such as stroke and hydrocephalus \cite{Mazurek2021,Obungoloch2018,Yuen2022,Laso2024}, and its demonstrated potential for portable whole-body screening \cite{Zhao2024,sheng2024breast,broche2024field} foreshadows a transformation in healthcare, particularly in remote and underserved areas \cite{webb2023five}.

Nevertheless, the clinical adoption of low-field MRI is hindered by its low signal-to-noise ratio (SNR) relative to clinical MRI, which is a direct result of the dependence of SNR on magnetic field  strength.\cite{Kimberly2023,Arnold2023} This low SNR reduces image clarity and hence limits diagnostic accuracy. To improve SNR,  scans are often repeated, and the data are averaged, but this comes at the cost of much longer scan duration, which can, in some cases, exceed one hour \cite{Sarracanie2015}. Such long scans limit practicality in clinical settings, particularly in critical care, and result in reduced patient comfort and compliance.

MRI scans can be accelerated through data acquisition with a sub-Nyquist rate, a process known as $k$-space (Fourier domain) undersampling. However, this introduces artifacts that are then removed with image-reconstruction algorithms that infer images from the undersampled data. Compressed Sensing (CS) \cite{Lustig2007,feng2017compressed} and deep learning (DL) \cite{wang2021deep,hammernik2023physics,Heckel2024} frameworks have been established as two powerful approaches that enable high-quality reconstructions from undersampled data. CS utilizes prior knowledge about image sparsity in certain transform domains, such as the wavelet domain, while DL methods learn priors directly from the underlying data. Both CS and DL approaches have had a transformative impact on high-field MRI \cite{wang2021deep,hammernik2023physics,Heckel2024}. 

Despite their success in high-field MRI, the application of CS and DL remains limited in the low-field MRI domain. CS methods have been explored only in a few studies \cite{Sarracanie2013compressed,Tamada2014,Koolstra2021}, e.g. in the context of Overhauser-enhanced MRI, where SNR is boosted by dynamic nuclear polarization techniques \cite{Sarracanie2013compressed}, or for the correction of image distortion \cite{Koolstra2021}. Meanwhile, DL has primarily been applied to image post-processing tasks such as de-noising \cite{Hernandez2021,Le2021deep,lin2024zero}, super-resolution \cite{Lau2023pushing,DeLeeuwdenBouter2022}, image-to-image translation \cite{Kim2023d,Islam2023,Lucas2023multi,Ayde2022,Oved2025}, and artifact detecting \cite{jimeno2022artifactid}, or for reducing the need for electromagnetic interference (EMI) shielding \cite{Liu2021,Zhao2024electromagnetic}. Very recently, a few studies have begun to explore using DL for image reconstruction at low field \cite{zhou2022dual,Man2023, Koonjoo2021}. However, the full potential of CS and DL frameworks for accelerating low-field MRI and enhancing its SNR has yet to be fully harnessed.

\hl{Because portable systems are often used outside shielded MRI suites, they are exposed to variable environmental electromagnetic interference, which can further degrade their already low intrinsic SNR \cite{Deoni2022}}. The performance of CS and DL methods has been shown to depend on noise variations in high field MRI \cite{antun2020instabilities, darestani2021measuring, knoll2019assessment,Shimron2022} but is relatively unexplored at low field. Additionally, reconstruction methods are typically evaluated using metrics that quantify the reconstruction quality relative to reference data. However, in the context of low-field imaging, this reference data can often be corrupted by substantial noise. While studies in high-field MRI have shown that using unsuitable reference data can lead to biased, overly optimistic results \cite{Shimron2022,wang2024hidden}, this critical matter has yet to be explored in the context of low-field imaging.

\begin{figure}
    \centering
    \includegraphics[width=86mm]{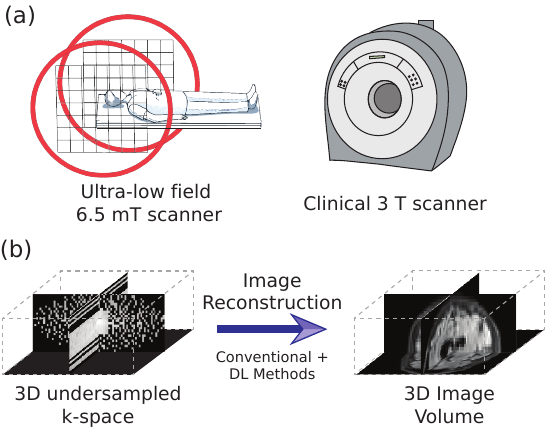}
    \caption{\textbf{Reconstruction accuracy across field strengths.}
    \textbf{(a)} MRI data in this study were collected using an ultra-low field (6.5~mT) MRI scanner based on a bi-planar electromagnet \cite{Sarracanie2015}, and a conventional 3~T MRI scanner, as illustrated schematically. Additional data were sourced from the fastMRI database and synthetically degraded by the addition of Gaussian noise to allow for a wide signal-to-noise ratio regime to be analyzed \cite{Knoll2020}.
    \textbf{(b)} Raw 3D $k$-space data were collected via retrospective and prospective acquisitions using a range of sampling patterns. This figure demonstrates a 3D variable-density undersampling pattern. The accuracy of 3D volumes reconstructed from $k$-space data using conventional methods such as compressed sensing was compared to leading data-driven and model-driven techniques based on neural networks.
    }
    \label{fig:overview}
\end{figure}

Here, we present a comprehensive study that addresses the research gaps in CS and DL reconstruction for accelerated low-field MRI. First, we explore approaches to enable rapid scans while also enhancing image quality. Specifically, we explore different strategies for distributing $k$-space samples within a fixed scan-time budget and reconstructing images, focusing on strategies combining $k$-space under-sampling with a variable number of excitations/repetitions (NEX) and CS reconstruction. Second, to further improve image quality and to ensure robustness across the SNR spectrum, we explore and compare methods from three well-established reconstruction frameworks - CS, data-driven \hl{DL}, and physics-guided DL methods - and study their performance for varying SNR levels. \hl{To explore how architectural design influences performance in physics-guided reconstruction, we compare convolution-based unrolled networks with attention-based Swin Transformer cascades, both of which incorporate data consistency constraints.} Next, to raise awareness to critical algorithmic evaluation issues, we conduct the first experiments demonstrating the `hidden noise problem' in low-field MRI. Finally, as illustrated in Figure \ref{fig:overview}, we evaluate image reconstruction methods using raw high-field data and prospectively undersampled data from an ultra-low field (6.5~mT) system. 

In summary, this work provides the first thorough investigation of MRI reconstruction frameworks across a \textit{broad SNR spectrum} that represents diverse low-field MRI settings. To the best of our knowledge, this aspect of low-field imaging has not been studied before. Specifically, it makes the following key contributions: 
(1) Exploration of strategies for maximizing speed and image quality within a fixed scan duration. The results show that a strategy combining $k$-space undersampling, scan repetition (NEX), and iterative reconstruction can outperform fully-sampled acquisitions and provide robustness to varying SNR.
\hl{(2) A comparative evaluation of state-of-the-art MRI reconstruction frameworks for low-field MRI. The results demonstrate that physics-guided DL methods—including unrolled convolutional networks and Swin Transformer cascades—consistently outperform CS and data-driven DL approaches. While the Swin Cascade achieved the highest reconstruction accuracy at high SNR and high field strength, unrolled networks were more robust at ultra-low SNR—highlighting that simpler, locally-aware architectures can outperform more complex models in challenging noise conditions.}  
(3) Experimental demonstration of the studied CS and DL frameworks using raw data from both high-field (3~T) and low-field (6.5~mT) scans, including experiments with prospectively undersampled data.  
(4) The first demonstration of the \textit{hidden-noise problem} in low-field MRI, which shows that computing image quality metrics relative to reference images containing different levels of noise can change the algorithmic ranking. \hl{As low-field MRI is inherently noisy, this demonstration contributes to growing awareness of the hidden noise problem and the challenges it poses for accurate algorithm assessment.}

\section{Methodology}

\subsection{Data Acquisition}

\subsubsection{Data preprocessing for \textit{in silico} experiments}

To generate large training, validation, and testing datasets of high-SNR images, brain data from the fastMRI database were preprocessed to resemble single-channel $k$-space data acquired on our 6.5~mT MRI scanner \cite{Sarracanie2015, Knoll2020}. This raw data was made publicly available by the NYU fastMRI Initiative specifically for the purpose of exploring whether machine learning can aid in the reconstruction of medical images \cite{Knoll2020,Zbontar2018}. In this experiment, multi-channel, fully-sampled raw $k$-space data were sourced from the fastMRI brain training and validation datasets, yielding thousands of 1.5~T / 3~T scans. Coil sensitivity maps were calculated using the ESPIRiT algorithm and SigPy \cite{Uecker2014,Ong2019} and used for combining the multi-channel inverse fast Fourier transform (IFFT) reconstructed images into complex-valued, single-channel images. Reconstructed brain volumes were converted to 3D $k$-space via fast Fourier transform (FFT) and downsampled to the ultra-low field resolution of $64 \times 75 \times 25$ via $k$-space cropping. Ground-truth reference images were obtained via IFFT reconstruction of these downsampled $k$-space data. Finally, the brain volumes were split into training, validation and test datasets containing 3300, 1100 and 100 volumes respectively.

The training dataset was further supplemented with 50,000 brain image slices sourced from the Human Connectome Project (HCP) \cite{van2013wu}. These brain images were derived from DICOM magnitude images and had synthetic phase maps applied to synthesize complex data acquisition. Individual image slices were downsampled to $75 \times 25$ via $k$-space cropping to match the resolution of ultra-low field acquisitions. The synthesis of this HCP dataset is fully described in Reference \citenum{Koonjoo2021}. We emphasize that this dataset was used only for training, i.e. it served as an augmentation method, and not for testing the algorithms.

To analyse the impact of SNR on image-reconstruction performance, for all datasets the downsampled $k$-space data were degraded through the addition of complex-valued additive white Gaussian noise at varying noise power. The quality of images reconstructed from this degraded $k$-space data was subsequently compared to reference images reconstructed from the original fully-sampled and noise-free data using the methods described below. As SNR scales approximately as $B_0^\frac{3}{2}$ \cite{Hoult1976}, we tested a wide range of SNR values (40~dB range shown in Fig. \ref{fig:fig2_fastmri_sampling} and Fig. \ref{fig:fig3_fastmri_metric_curves}) as the variation in SNR expected between clinical and low-field MRI scanners can be very large. Additionally, images and $k$-space data were linearly scaled such that the 95th percentile of pixel intensity in IFFT reconstructed images was set to 1.

\subsection{MRI Data Acquisition}
Experimental data in this study were acquired at the ultra-low field of 6.5~mT and the standard clinical field strength of 3~T. All data were acquired with informed consent in compliance with Institutional Review Board (IRB) approvals. MRI data were acquired under Mass General Brigham protocol number 2022P002344, which was approved on July 10, 2024.

\subsubsection{Ultra-low Field MRI Experiments}
Imaging experiments were performed on a 6.5~mT MRI scanner that consists of a biplanar electromagnet-based scanner with a peak gradient strength of 1~mT/m \cite{Sarracanie2015}. Image data were acquired with a custom quadrature head coil that has been described previously \cite{Koonjoo2017}. Data from this quadrature coil were digitized as two-channel data and then linearly summed after applying a $\pi/2$ phase shift to $k$-space data from the second channel. \hl{The 6.5~mT MRI system includes an RF-shielded enclosure, as described in Reference \citenum{Sarracanie2015}, which reduces environmental electromagnetic interference during acquisition.}

A 3D Cartesian balanced-steady state free precession (bSSFP) sequence with TR/TE = 22/11 ms and matrix size $64 \times 75 \times 25$ (Readout $\times$ Phase Encode 1 $\times$ Phase Encode 2) was programmed with a Tecmag Restone spectrometer and used for data acquisition. These parameters gave a voxel size on the 6.5~mT scanner of 2.5~mm$\times$3.5~mm$\times$8~mm (RO$\times$PE1$\times$PE2). The number of excitations (NEX) was varied as described in the Results section. \hl{A bSSFP sequence is utilized to maximize SNR per unit time, which is critical in the ultralow field domain \cite{Sarracanie2015}.}

For prospectively undersampled acquisitions, Poisson disc masks with acceleration factors of $R=2$ and $R=4$ were exported from SigPy and used to populate the phase encoding tables on the spectrometer prior to acquisition. Fully-sampled acquisitions were captured in the same imaging session for comparison of reconstruction methods. All scans were acquired independently with $\mathrm{NEX}=48$, so that the accelerated acquisitions were completed in half or a quarter of the fully-sampled time for $R=2$ and $R=4$ masks, respectively.

\subsubsection{MRI at 3T}
High-field imaging data were collected in compliance with IRB approvals from a healthy volunteer on a 3~T Siemens Skyra scanner. A single-channel head coil and a 3D bSSFP sequence were used for data acquisition. The 3~T bSSFP data shown in Figure \ref{fig:fig6_retro_3T_ulf} were acquired with TR/TE~=~4.8~ms/2.4~ms and a matrix size of $256 \times 256 \times 128$ (Readout $\times$ Phase Encode 1 $\times$ Phase Encode 2) in the anterior-posterior, left-right and superior-inferior directions, respectively. Fully-sampled data were retrospectively downsampled via $k$-space cropping to the same resolution used at ultra-low field ($64 \times 75 \times 25$). Fully-sampled data were additionally acquired with gradient-recalled echo (GRE) and Magnetization Prepared Rapid Gradient Echo (MPRAGE) sequences using TR/TE values of 8~ms/2.3~ms and 2300~ms/4.4~ms, respectively.

\subsection{Image Reconstruction}

\subsubsection{Retrospective undersampling}
For retrospective undersampling experiments, 2D masks were created in Python using Sigpy \cite{Ong2019} and then applied to fully-sampled data in both phase encode directions. Poisson-disc masks were created using the `Poisson-disc' tool in Sigpy, with a $10 \times 10$ fully-sampled region added to the center. The reported acceleration factors (2 or 4) were calculated after addition of this fully-sampled region. Additional masks used for undersampling are shown with the accompanying results where relevant.

\subsubsection{IFFT and CS reconstructions}
Fully-sampled data were reconstructed using a conventional IFFT operation. For undersampled data, missing $k$-space data entries were first zero-filled, and images were then reconstructed via IFFT. The 95th percentile intensity value of the zero-filled images was linearly scaled to 1 and the resultant scaling factor applied to the undersampled $k$-space as a starting point for subsequent $l1$-wavelet CS reconstruction using the Sigpy package \cite{Ong2019}. The CS regularization parameter ($\lambda$) was optimized for each dataset and acceleration factor separately by performing a grid search over the range of [10$^{-6}$-10$^{-1}$]. The iteration number used for the reconstructions was then optimized over the range [1-200].

For experiments using the fastMRI dataset, the optimal $\lambda$ and iteration numbers for CS reconstruction were calculated as the values that minimized the mean normalized root-mean-square error (NRMSE) across 20 brain volumes drawn from the validation set as compared to the ground truth image. Values were re-optimized for each variation in input SNR. For retrospective and prospective experiments on our MRI scanners, the $\lambda$ and iteration values were chosen to minimize NRMSE relative to the fully-sampled image. This parameter re-optimization means that the CS algorithm is being evaluated in terms of its `best possible performance' in each experiment.

\subsubsection{DL reconstruction}
We analyzed the performance of \hl{three} DL reconstruction frameworks: (i) a data-driven framework trained on the forward encoding model known as AUTOMAP \cite{Zhu2018}, (ii) a physics-guided unrolled neural network \cite{Zhang2018,Shan2023} \hl{and (iii) a Swin Transformer cascade.} 


\textbf{\textit{AUTOMAP:}} The AUTOMAP reconstruction framework was implemented using supervised learning in Keras and TensorFlow \cite{Koonjoo2021,Waddington2023}. The AUTOMAP network architecture comprises six trainable layers. The first two layers are fully-connected (dense) layers with hyperbolic tangent activations, which process the flattened input data through hidden layers. The data are then reshaped into a matrix format before passing through three convolutional layers. The first convolutional layer has 128 filters with a 5 × 5 kernel and uses a hyperbolic tangent activation function. The second and third convolutional layers each have 128 filters with 5 × 5 kernels and utilize rectified linear unit (ReLU) activation functions. Finally, a transposed convolutional layer with a single filter and a 5 × 5 kernel generates the final image. This architecture enables the network to learn the spatial decoding process between $k$-space and image space without any prior knowledge. The AUTOMAP network contains 21,925,360 trainable parameters for the $75 \times 25$ input size. \hl{TensorFlow was utilised to ensure that the AUTOMAP architecture was implemented in the same fashion as originally published in Reference \citenum {Zhu2018}.} Visual descriptions of the AUTOMAP architecture were provided in References \citenum{Zhu2018}, \citenum{Koonjoo2021} and \citenum{Waddington2023}.

\textbf{\textit{Unrolled DL:}} Our unrolled DL approach to image reconstruction was implemented in PyTorch. The Unrolled architecture consists of 10 iterative soft shrinkage-thresholding blocks. Each block begins with a data fidelity calculation, followed by nonlinear transforms and a soft thresholding operation. The nonlinear transforms include forward and backward operations, each designed as a combination of two convolutional layers split by a rectified linear unit (ReLU). A skip connection is incorporated to form a residual block, aiding in network training. A data consistency (DC) operation is included at the end of each block (except for the final block) to ensure $k$-space consistency with the measured $k$-space samples. This DC layer replaces DL-generated $k$-space data with the original data at sampled points. In the last iteration, the DC block is omitted to allow the convolutional layers to de-noise the original low SNR $k$-space data that was re-injected in previous iterations. As implemented, the Unrolled network has  98,760 parameters, a value that is not dependent on the input size. The full Unrolled network architecture is detailed in Reference \citenum{Shan2023}.

\hl{\textbf{\textit{Swin Cascade:}} The Swin Transformer cascade was implemented in PyTorch using transformer blocks adapted from the SwinIR (Swin Transformer for Image Restoration) framework \cite{liang2021swinir, Rahman2025}. The model comprises six cascaded stages, each operating in the image domain and based on residual Swin Transformer layers with \textit{shifted window self-attention}. Each stage refines the reconstruction using these SwinIR-based transformer blocks. The output is passed through an FFT to update $k$-space before DC. The DC operation was implemented by replacing predicted $k$-space values at sampled locations with the original acquired measurements. This architecture enables the network to process data efficiently with physics constraints, while enabling full denoising in the last stage. The total number of trainable parameters was 8,335,032 for the training input size used. Although this structure mirrors the image–data-consistency staging of our convolutional unrolled DL network, we follow standard terminology and keep the transformer variant labelled a \textit{Swin Cascade}—similar to other work that explored transformers for high-field MRI \cite{Rahman2025,Sheng2024}—while reserving the term \textit{unrolled} for architectures that explicitly unfold a specific iterative optimizer using convolutional layers. Additional details of the Swin Transformer design and its application in a Swin Cascade are provided in References \citenum{liang2021swinir} and \citenum{Rahman2025}.}

\textbf{\textit{Training:}} \hl{All} image reconstruction networks were trained using the same training corpus of fastMRI and HCP data described above, which consisted of more than 250,000 image slices. To minimize graphical processing unit (GPU) memory requirements, networks were trained to reconstruct 2D, rather than 3D, data \cite{Eo2021}. Hence, the training corpus was utilized as a stack of independent 2D $75 \times 25$ brain image/masked $k$-space pairs. Data were further augmented by adding random amounts of synthetic Gaussian noise to the input during training, with a variable magnitude range that mimicked the SNR expected during implementation. Individual networks were trained for each undersampling mask. \hl{Hyperparameters such as learning rate, and number of unrolls or cascades were selected via manual tuning on a validation subset, balancing training stability, reconstruction fidelity, and GPU memory constraints.}

A custom loss function was defined to calculate the mean square error (MSE) after masking the background region outside the brain/head. This masking encourages the neural networks to focus on improving image structure within the brain rather than de-noising the background \cite{Sun2019}. Masking was performed in regions where the normalized reference images had a signal magnitude less than 0.005. \hl{All networks were trained for up to 100 epochs or 100 hours, whichever came first. While the Unrolled and AUTOMAP models were trained for the full 100 epochs, the Swin Cascade models were trained for 12 epochs, which was sufficient for convergence in all cases. Although the Swin Cascade has \textasciitilde8.3\,M learnable parameters---fewer than the \textasciitilde20\,M in AUTOMAP---it stacks many more layers and performs multi-head windowed self-attention in every block. As a result, the Swin model uses substantially more GPU memory and takes longer to train than AUTOMAP, despite its smaller parameter count \cite{kaplan2020scaling,liu2021swin}.} AUTOMAP training utilized a learning rate of 0.0001 and the RMSprop optimizer. Unrolled DL networks were trained with a learning rate of 0.001 and the Adam optimizer. \hl{Swin Cascade networks were trained with a learning rate of 0.0001 and the Adam optimizer.} Two separate AUTOMAP networks were trained for each reconstruction task, one to reconstruct real image data and one to reconstruct imaginary image data. AUTOMAP networks received the same complex $k$-space input data but were shown different (real or imaginary) reference images.

Additional experiments were performed to test the effects of perceptual loss functions during training \cite{Zhang2018lpips}. To train the unrolled network, we utilized the Learned Perceptual Image Patch Similarity (LPIPS) loss available for PyTorch at \url{https://github.com/richzhang/PerceptualSimilarity}. To train AUTOMAP, we ported this LPIPS  loss to TensorFlow. Our implementation is available at: \newline \url{https://github.com/Image-X-Institute/lpips_torch2tf} \cite{Grover2024lpips}.

\textbf{\textit{Inference.}}
All reconstructions of test data were performed on 3D $k$-space data. To address GPU memory limitations, a 1D FFT was first applied along the fully-sampled readout dimension of $k$-space to yield a stack of hybrid $k$-space data. The trained networks were then applied to the hybrid $k$-space data to perform reconstruction in undersampled phase encode directions via inference. In the case of AUTOMAP, inference was performed separately with real and imaginary networks before outputs were combined into a complex image.

\subsubsection{Metric Calculations and Hidden Noise}
The quality of brain volumes reconstructed using different sampling approaches was quantitatively evaluated using normalized root-mean-square error (NRMSE) and structural similarity (SSIM) metrics. Maximum-minimum normalized root-mean-square error (NRMSE), and structural similarity index (SSIM) metrics were calculated relative to a fully-sampled reference image \cite{Wang2004}. \hl{Structural similarity index (SSIM) was computed on a slice-by-slice basis across the 3D volume, and the reported SSIM values represent the mean of these slice-wise scores. This approach, commonly referred to as \textit{slice-wise SSIM}, enables robust comparison across volumes of varying anatomy and SNR.} Regions outside the brain/phantom that should not contain any MRI signal were excluded from metric calculations by calculating a binary mask on the fully-sampled reference image. By excluding the background from metric calculations, algorithms that better recover structures within the region of interest are ranked more highly. For \textit{in silico} experiments utilizing fastMRI data, the fully-sampled reference is the original image with no noise applied.

In our experiments exploring the impact of hidden noise on error metrics, fully-sampled reference data were acquired from a head-shaped phantom at 6.5~mT with a NEX of 16 and 256. Error metrics for images reconstructed from undersampled data ($R=2$, NEX~=~16) were calculated against each reference independently.

\section{Results}


\subsection{Signal averaging and subsampling for efficient acquisitions at low SNR}

\begin{figure*}
    \centering
    \includegraphics[width=1\linewidth]{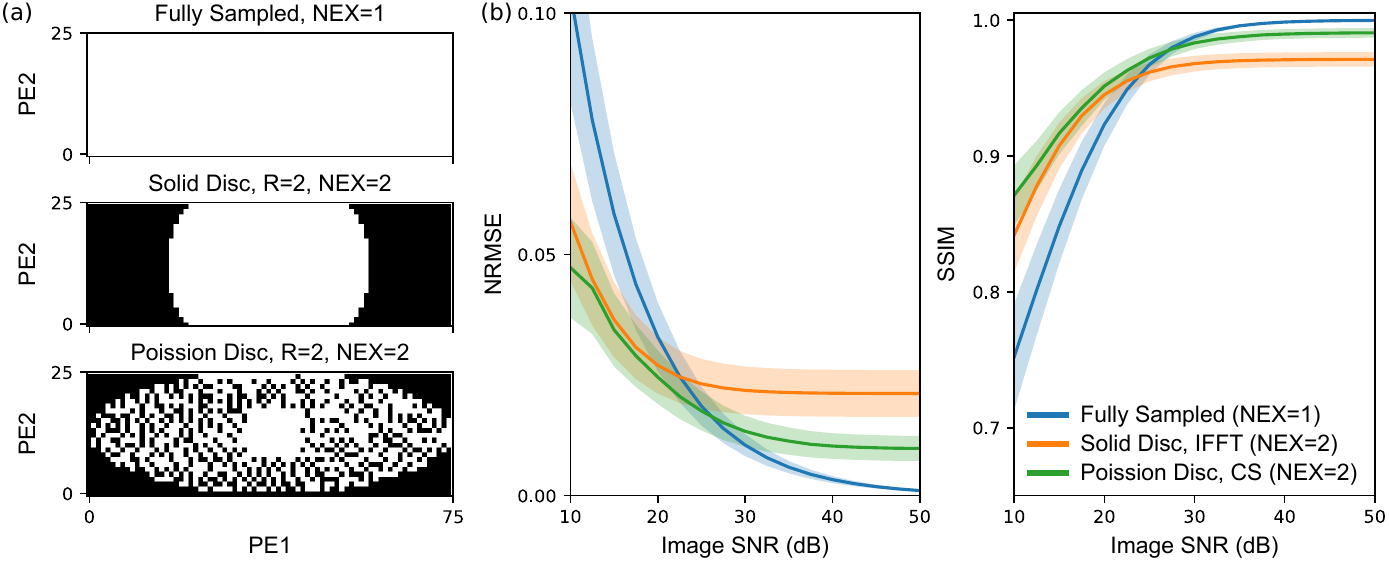}
    \caption{\textbf{Comparison of three sampling-reconstruction strategies for utilizing a fixed total scan time `budget', with evaluation across varying SNR values.} 
    \textbf{(a)} Illustration of the three strategies. (1) Acquisition of fully-sampled $k$-space data without repetitions (NEX~=~1) and inverse Fast Fourier Transform (IFFT) reconstruction. (2) Solid disc under-sampling with $R=2$ and two repetitions (NEX~=~2) and compressed sensing (CS) reconstruction. (3) Poisson disc undersampling with $R=2$ and two repetitions (NEX~=~2) and CS reconstruction. \textbf{(b)} Evaluation of the three strategies using input data with various SNR levels. Reconstruction quality was measured using the normalized root-mean-square error (NRMSE) and structural similarity index (SSIM) metrics. Solid lines indicate the mean and shaded regions indicate the standard deviation across 100 brain volumes. The $x$-axis SNR values are calibrated to the Image SNR of the fully-sampled image (i.e. 20~dB is a linear Image SNR of 10, 40~dB is a linear Image SNR of 100). Notably, the results indicate that the combination of 2$\times$ Poisson Disc undersampling, 2 repetitions and CS reconstruction results in higher reconstruction quality than the other two approaches at low SNR values.  
    }
    \label{fig:fig2_fastmri_sampling}
\end{figure*}

Data acquisition with low- and especially ultra-low field MRI systems often involves many repetitions per $k$-space line to improve SNR. Results from our first experiment, shown in Figure \ref{fig:fig2_fastmri_sampling}, characterize the potential of undersampling and CS methods to improve image quality without changing the overall scan duration. Reconstruction quality metrics are shown for different sampling approaches that distribute samples within a fixed scan time `budget', by reducing the number of points acquired and increasing NEX. Specifically, metrics calculated on the fastMRI test set are shown for three sampling strategies (a): (1) a fully-sampled acquisition without repetitions (i.e. NEX~=~1) and IFFT reconstruction; (2) a 2$\times$ accelerated scan ($R~=~2$) where data are undersampled using a \textit{solid disc} mask or (3) a \textit{Poisson disc} mask. In cases (2) and (3) the overall scan time was identical to the first case, yet the data were undersampled, hence images were reconstructed via CS.

 Our results indicate that for very low SNR values (<25~dB), typical of low-field MRI, CS reconstruction of Poisson-disc undersampled data gives the best image reconstruction metrics (low NRMSE, high SSIM). In fact, an approach combining 2x undersampling, 2 repetitions, and CS reconstruction is superior to a single acquisition and IFFT reconstruction for the same acquisition time. This improvement at low SNR likely stems from the $\sim$3~dB increase in $k$-space SNR that comes with an additional signal averaging and the denoising effects of CS reconstruction. Conversely, at high SNR values ($>$30~dB), image quality metrics indicate that the best reconstructions are obtained by acquiring fully-sampled $k$-space data without repetitions.

In Supplementary Figure 1 we show results repeating our sampling experiment with an $R~=~2$ Poisson disc mask that fills the corners of $k$-space. Such changes in sampling distribution appear to have a limited impact on the quantitative reconstruction results. A thorough examination of even more quasi-random undersampling masks in ultra-low field MRI can be found in Reference \citenum{Waddington2022ismrm}. 


\subsection{Exploring DL-based reconstruction across the SNR regime}

\begin{figure}
    \centering
    \includegraphics[width=88mm]{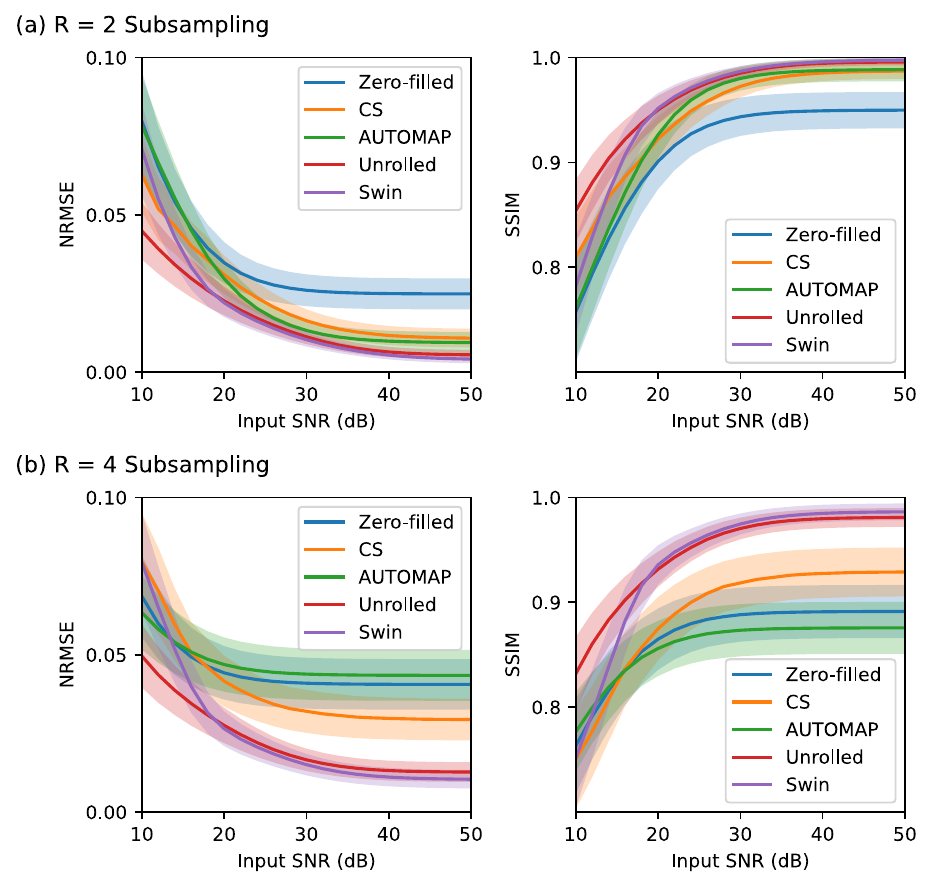}
    \caption{\textbf{Quantitative evaluation of reconstruction quality at different signal-to-noise ratios (SNR) levels.}
    The mean normalized root mean square error (NRMSE) and structural similarity (SSIM) of 100 brain volumes reconstructed from undersampled, low-resolution FastMRI $k$-space data are shown for acceleration factors ($R$) of 2 and 4 in \textbf{a} and \textbf{b}, respectively. The SNR of the input signal ($k$-space) was adjusted by adding synthetic Gaussian noise scaled to the ground truth $k$-space data. Four reconstruction methods were compared: zero-filled Inverse Fast Fourier Transform (blue), Compressed Sensing (CS, orange), AUTOMAP (green), Unrolled DL (red), and \protect\hl{Swin Cascade (purple)} techniques. Shaded regions indicate the standard deviation of the metrics. Note that the unrolled DL outperforms all other methods at \protect\hl{very low SNR} (lowest NRMSE, highest SSIM).
    }
    \label{fig:fig3_fastmri_metric_curves}
\end{figure}

\begin{figure*}
    \centering
    \includegraphics[width=0.9\linewidth]{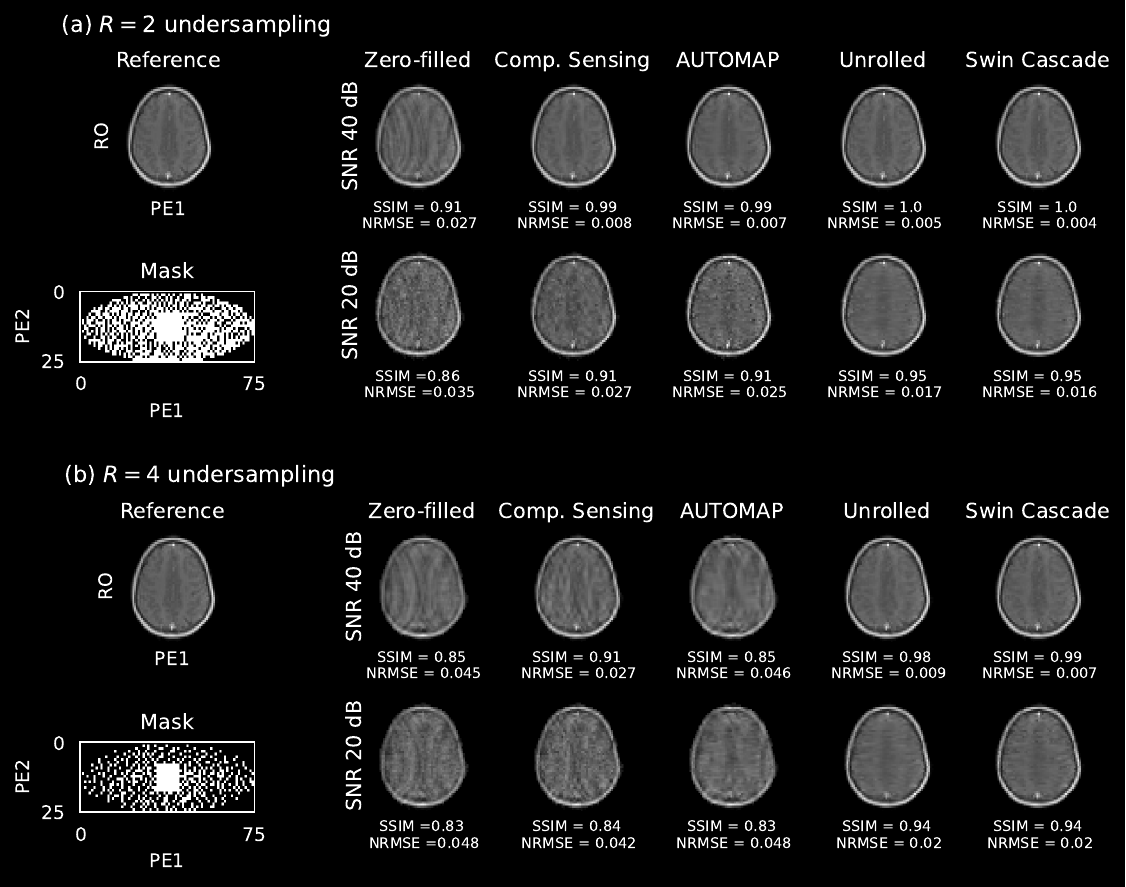}
    \caption{\textbf{Visual comparison of reconstruction performance on retrospectively undersampled FastMRI data.}
    Reconstructions of the central slice from a brain volume are shown for the zero-filled Inverse Fast Fourier Transform, Compressed Sensing, AUTOMAP, Unrolled DL and \protect\hl{Swin Cascade} methods, for acceleration factors ($R$) of 2 and 4 in \textbf{a} and \textbf{b}, respectively. The Poisson disc masks used for undersampling are also displayed, in a plane consisting of the Phase Encode 1 (PE1) and Phase Encode 2 (PE2) directions. The reconstructions are shown for $k$-space data with `high' and `low' signal-to-noise ratios (SNR) of 40~dB and 20~dB, respectively, in the PE1 and Readout (RO) plane. Note that for $R=4$, only the \protect\hl{physics-guided DL methods} can remove the undersampling artifacts.
    }
    \label{fig:fig4_fastmri_imgs}
\end{figure*}

Having established that Poisson disc undersampling can improve image quality when SNR is low, we now examine the comparative advantages of different reconstruction frameworks across SNR regimes. Metrics comparing CS and DL reconstruction of undersampled data are shown in Figure \ref{fig:fig3_fastmri_metric_curves}. \hl{Our findings indicate that both the unrolled network and the Swin Cascade provide the best reconstruction quality across the range of SNR values and acceleration factors analyzed, as quantified by the image quality metrics (i.e. low NRMSE, high SSIM).} \hl{The Swin Cascade achieves the highest performance at high SNR, while the unrolled network performs better in very low SNR regimes.} The visual comparison (Figure \ref{fig:fig4_fastmri_imgs}) further demonstrates that \hl{both methods produce high-quality reconstructions with minimal artifacts and well-preserved structural detail}. \hl{The performance difference between the two networks is small across most regimes, suggesting both architectures are highly effective for image reconstruction.} These results indicate that \hl{physics-guided DL methods outperform the other reconstruction strategies (IFFT, CS, and AUTOMAP) across field strengths and noise levels.}

Furthermore, we also observe that the relative ranking of CS and AUTOMAP is inconsistent; it depends on both the acceleration factor and the SNR level (Figure \ref{fig:fig3_fastmri_metric_curves}a). For example, for $R=2$, the quantitative metrics indicate that AUTOMAP outperforms CS reconstruction in the high SNR regime but that when SNR is reduced ($<$19~dB), CS performs better. However, the reconstructions of both methods exhibit similar quality, without noticeable aliasing artifacts (Fig. \ref{fig:fig4_fastmri_imgs}a). In contrast, at $R=4$ (Fig. \ref{fig:fig4_fastmri_imgs}b), both AUTOMAP and CS yield images with aliasing artifacts, which are blurrier than the \hl{Unrolled and Swin Cascade network reconstructions}. While the noise in the AUTOMAP and CS reconstructions is observed to decline at high SNR (40~dB), the aliasing artifacts remain. This is also reflected in the NRMSE and SSIM quality metrics (Fig. \ref{fig:fig3_fastmri_metric_curves}b). Importantly, at $R=4$, the standard deviation of NRMSE/SSIM metrics is observed to decrease with the SNR level for the \hl{Unrolled and Swin Cascade} networks but not for CS and AUTOMAP, indicating improved consistency and robustness of the physics-guided DL methods. \hl{These results clarify that the similarity between AUTOMAP and Unrolled Networks is restricted to low acceleration and high-SNR conditions; at $R=4$, the unrolled network clearly outperforms AUTOMAP in both quantitative metrics and visual detail.}

In summary, the results show that the \hl{Unrolled and Swin Cascade} networks outperform other methods and exhibits high robustness to SNR levels. The performance of CS and AUTOMAP is not consistent when the acceleration factor increases from $R=2$ to $R=4$, and \textit{only} the physics-guided DL networks can effectively recover fine details for $R=4$.

\subsection{The hidden noise problem in ultra-low field MRI}

The quantitative comparisons we have shown thus far have made the implicit assumption that the reference data, which is used for quantifying algorithmic performance, consists of noiseless `ground truth' images. However, in MRI, it is infeasible to experimentally measure a noiseless image. \hl{Moreover, it is increasingly recognized that even small variations in the reference image can distort quantitative metrics such as NRMSE and SSIM, especially when reference data contains residual noise or is acquired with fewer averages.} \hl{This phenomenon—referred to as the `hidden noise problem'—is distinct from general criticisms of full-reference metrics. It concerns the scenario where the reference image itself is imperfect due to noise, which can penalize denoising algorithms or reorder the apparent ranking of methods, even when computed using a fixed metric.} \hl{While this issue has been studied in high-field MRI \cite{Shimron2022, wang2024hidden}, it has not yet been evaluated in the ultra-low field regime, where low SNR and long scan durations make high-quality references particularly difficult to acquire.}

\begin{figure*}
    \centering
    \includegraphics[width=1\linewidth]{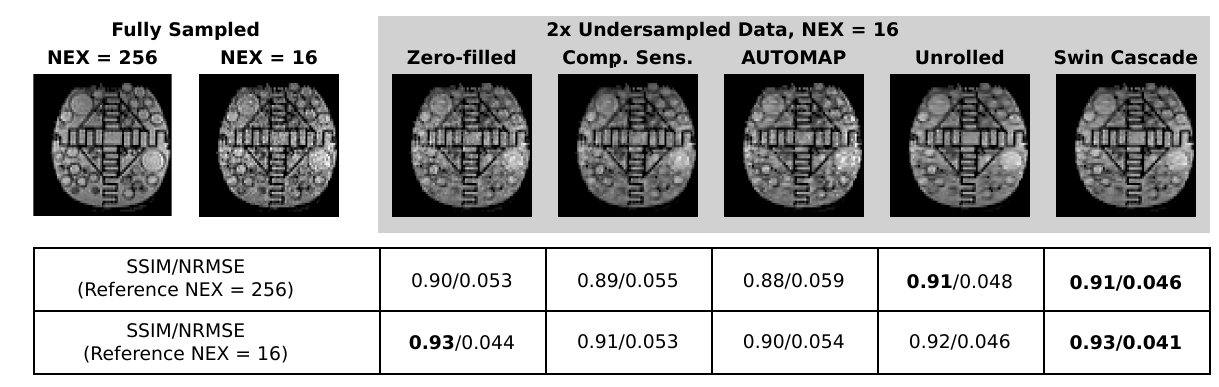}
    \caption{
    \textbf{The hidden noise problem in low-field MRI: computing image quality metrics (SSIM, NRMSE) using reference images with different levels of noise levels provides highly different results and changes the algorithmic ranking.}  The experiment was done using a 6.5~mT scanner and a 3D phantom; a middle slice from the phantom is displayed. Two different reference images were created by acquiring two fully-sampled datasets where the number of excitations (NEX) was 16 and 256.  Then, images were reconstructed using the IFFT. Notice that the NEX=16 image contains much more noise than the NEX=256 one. To evaluate the reconstruction methods, the NEX=16 data were retrospectively undersampled with an $R=2$ Poisson-disc mask and reconstructed with four methods: zero-filled inverse fast Fourier transform (IFFT), Compressed Sensing, AUTOMAP, and the unrolled network. The SSIM and NRMSE metrics were calculated relative to the two fully-sampled reference images, for all the pixels in the 3D volume. The table shows the results for both cases. Notice how the ranking of the `best' result (highlighted in bold) is strongly influenced by the reference image choice.} 
    \label{fig:fig5_reference-noise}
\end{figure*}

To demonstrate this problem, we computed image quality metrics relative to two 6.5~mT reference scans with differing NEX values as shown in Figure \ref{fig:fig5_reference-noise}. \hl{Among all methods, the Swin Cascade achieves the best SSIM and NRMSE scores when evaluated against the high-SNR (NEX~=~256) reference, consistent with its high pixel-level fidelity.} Structures in the Unrolled DL reconstruction also have clear edges and are de-noised relative to other methods. However, when metrics are recalculated against the NEX~=~16 reference image, \hl{the ranking of Unrolled and IFFT reverses — the zero-filled IFFT now appears to outperform Unrolled DL, despite its visibly blurrier and less accurate reconstruction.} This effect arises because the same underlying $k$-space data were used for both the undersampled and reference images, and the IFFT operation does not modify the noise structure. \hl{AI-based methods, including Unrolled and Swin Cascade, perform denoising, which introduces mismatch when compared against a noisy reference, artificially lowering SSIM and increasing NRMSE.} \hl{Interestingly, while Swin Cascade remains top-ranked in both comparisons, its output appears slightly noisier than the Unrolled result. This may reflect the Swin model's capacity to retain or model long-range noise textures via attention, whereas the convolutional Unrolled network favors edge enhancement and local smoothing — better aligning with the phantom's ground truth of smooth structures and sharp boundaries.} Importantly, we note that using the same underlying $k$-space for reference and undersampled data is common practice in accelerated MRI studies due to the time challenges inherent in collecting a high SNR reference image; our NEX~=~256 scan took nearly 2 hours to acquire, a time-frame more amenable to motion-free phantom imaging than humans \cite{Knoll2020a}.

The DL networks used thus far were trained with mean-square error (MSE) loss. However, it is increasingly recognized that such `full reference' metrics are not the best metric for measuring image quality as perceived by humans \cite{Reinke2024}. Hence, we explored the impact of training the Unrolled DL network with a `perceptual loss' based on layer activations in the AlexNet CNN \cite{Zhang2018lpips}. While images reconstructed with a perceptually trained network showed poorer NRMSE and SSIM values, it appears that higher frequency details may be better reconstructed (see Supplementary Figure 2).

\subsection{Comparison for retrospectively undersampled raw data from high- and low-field MRI Scans}
\begin{figure*}
    \centering
    \includegraphics[width=1\linewidth]{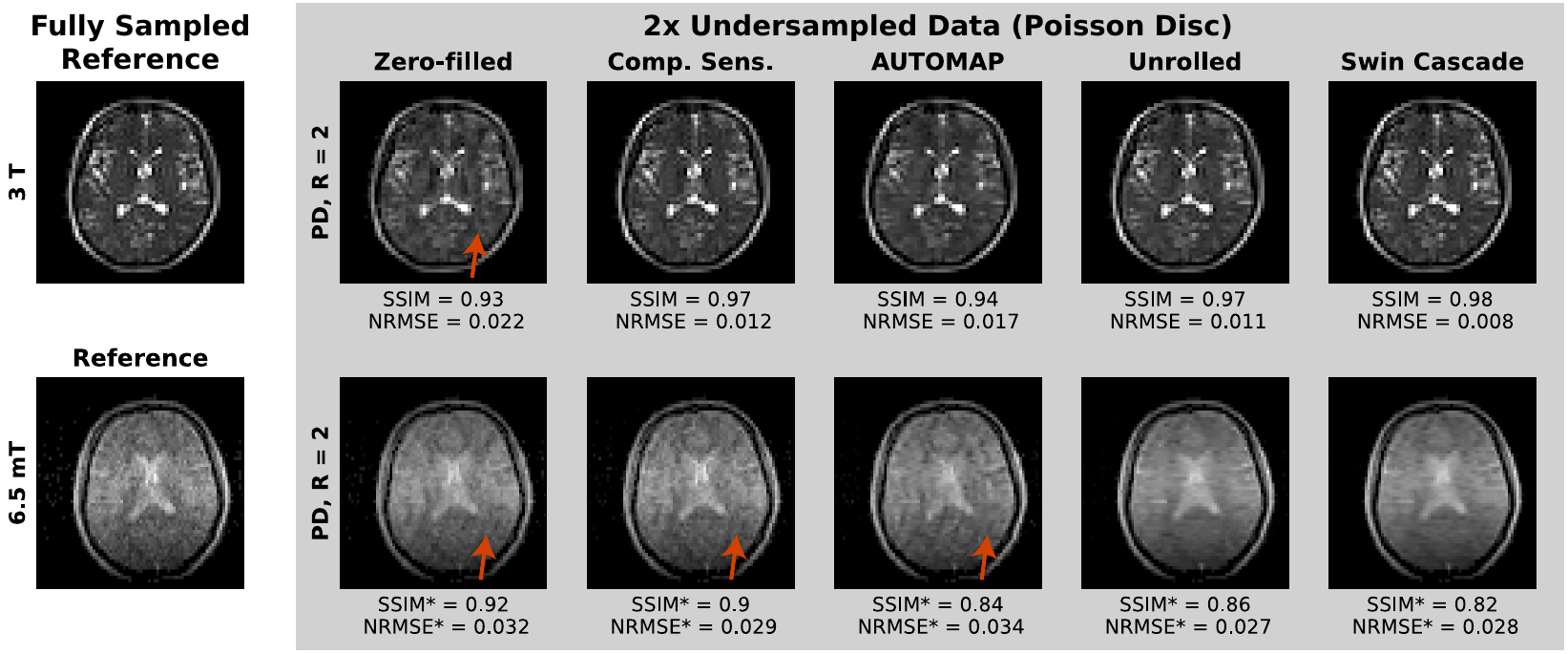}
    \caption{\textbf{Experimental comparison of reconstruction methods at high and low magnetic fields.}
    Fully-sampled, single-channel data from the same healthy subject were acquired at 3 T and 6.5 mT. A Poisson Disc mask ($R=2$) was retrospectively applied to the $k$-space data. Compressed sensing (CS), AUTOMAP, Unrolled DL and \protect\hl{Swin Cascade} reconstructions were effective at 3 T. In contrast, only the Unrolled DL and \protect\hl{Swin Cascade} removed undersampling artifacts (red arrows) at 6.5 mT. Normalized root-mean-square error (NRMSE) and Structural similarity (SSIM) values are starred (*) for 6.5~mT images to indicate that the values should be interpreted with caution due to noise present in the fully-sampled reference.}
    \label{fig:fig6_retro_3T_ulf}
\end{figure*}

Mindful that the repurposing of publicly available datasets (e.g., fastMRI) for testing DL methods can lead to overly optimistic results \cite{Shimron2022}, we next evaluated the performance of our reconstruction methods on raw $k$-space data from 3D human brain scans acquired at 6.5~mT and 3~T MRI scanners (Figure \ref{fig:fig6_retro_3T_ulf}). We find that CS, AUTOMAP, \hl{Unrolled DL, and Swin Cascade} produce visually similar images for the 3~T data, all free of undersampling artifacts and noise. In contrast, for the 6.5~mT data, the zero-filled, CS, and AUTOMAP methods yield reconstructions with visible noise and residual artifacts (see arrows in Figure \ref{fig:fig6_retro_3T_ulf}), indicating that these approaches are less robust under low-SNR conditions. The \hl{Unrolled DL and Swin Cascade} methods, on the other hand, provide denoised, artifact-free images with clear structural detail. \hl{Despite being significantly lighter in terms of parameter count and architectural complexity, the Unrolled DL model achieves the lowest NRMSE and outperforms all other methods on this dataset.} In contrast, the SSIM metric is highest for the zero-filled IFFT reconstruction — despite its visible degradation — a discrepancy likely attributable to hidden noise in the fully-sampled reference, as discussed in the previous section. This experiment highlights the importance of testing reconstruction methods on real-world MRI data, where noise and scanner variability can expose limitations not evident in curated datasets like fastMRI (Figure \ref{fig:fig4_fastmri_imgs}). \hl{It also reinforces the need to interpret image quality metrics cautiously in the presence of reference noise, particularly at ultra-low field.}

Significant differences in image contrast are seen between the bSSFP brain images acquired at 6.5~mT and 3~T (Fig. \ref{fig:fig6_retro_3T_ulf}). These contrast differences largely stem from the increase in $T_2$ relaxation times observed in the ultra-low field regime and the $T_2/T_1$-weighting of bSSFP sequences \cite{Waddington2020}. Likely, the reduced performance of the data-driven AUTOMAP reconstruction method has resulted from a shift between training and testing domains, which can be caused by changes in image contrast \cite{Waddington2023}. Model-driven DL methods, such as our Unrolled DL and \hl{Swin Cascade} approaches, are frequently found to be more robust to these domain shifts \cite{Shan2023}. In Supplementary Figure 3, we also show that the \hl{physics-guided DL} methods robustly reconstruct data acquired with gradient echo (GRE) and Magnetization Prepared Rapid Gradient Echo Imaging (MPRAGE) sequences at 3~T--these sequences are pillars of clinical imaging.

\subsection{Prospective undersampling at ultra-low field}

\begin{figure*}
    \centering
    \includegraphics[width=1\linewidth]{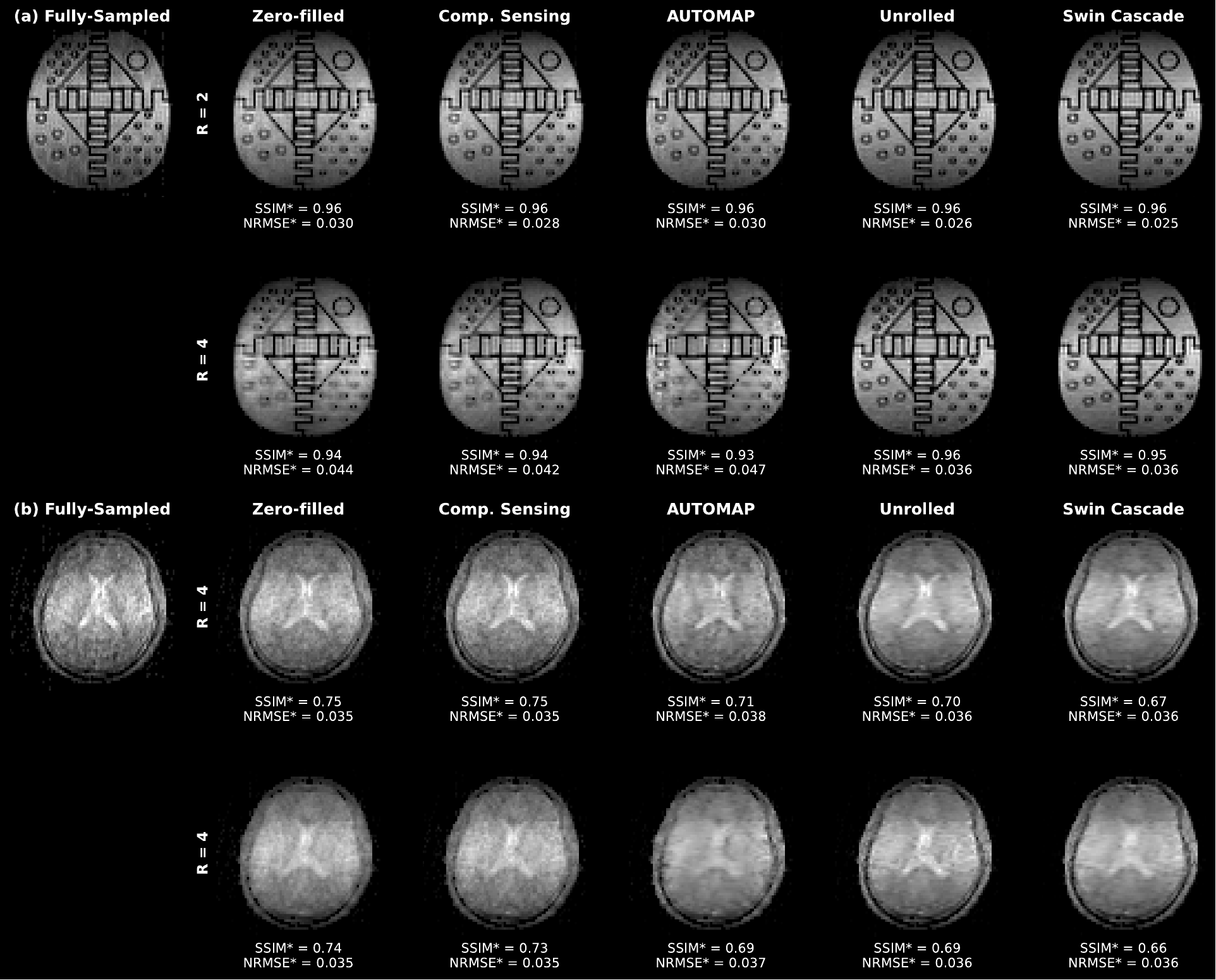}
    \caption{\textbf{Reconstruction of prospectively undersampled, raw $k$-space data acquired with ultra-low field MRI.}
    Poisson disc masks with $R=2$ and $R=4$ undersampling factors were deployed to a 6.5~mT MRI scanner and used to image a phantom and a human brain, as shown in \textbf{a} and \textbf{b} respectively. Normalized root-mean-square error (NRMSE) and structural similarity (SSIM) were calculated relative to fully-sampled images that were independently acquired with the same number of excitations (NEX). Phantom scans (NEX=48) took 34~min., 17~min. and 8.5~min for fully-sampled, $R=2$ and $R=4$ scans, respectively. Brain scans (NEX=64) took 44~min., 22~min. and 11~min. for fully-sampled, $R=2$ and $R=4$ scans, respectively. Image slices from volumes reconstructed with Zero-filled Inverse Fast Fourier Transform, Compressed Sensing, AUTOMAP, \protect\hl{Unrolled DL and Swin Cascade} techniques are shown. Quantitative image quality metrics are starred (*) to indicate that the values should be interpreted with caution due to hidden noise present in the fully-sampled reference. Note that the unrolled DL produces the sharpest results in all cases.}
    \label{fig:fig7_prospective-deployment}
\end{figure*}

Our final experiment evaluates the performance of our reconstruction methods on \textit{prospectively undersampled} ultra-low field data acquired in phantom and in-vivo brain scans using our 6.5~mT MRI scanner (results shown in Figure \ref{fig:fig7_prospective-deployment}). 

Phantom images (Fig. \ref{fig:fig7_prospective-deployment}a) are consistent with the results presented thus far. The Unrolled DL method produced the highest-quality images, with sharper edges and lower noise than the other methods. Specifically, for $R=4$, the CS and AUTOMAP methods yielded blurry images. \hl{The Swin Cascade also performed well, and the differences between Swin and Unrolled in this phantom setting were negligible, both visually and in terms of reconstruction metrics.} 

\hl{The prospectively undersampled in vivo brain images (Fig. \ref{fig:fig7_prospective-deployment}b) reveal clear anatomical structures such as the dura and ventricles, which appear sharpest in the Unrolled DL reconstruction.} Both the physics-guided DL and AUTOMAP methods reduced noise compared to CS, yet the physics-guided models did not blur the image to the same extent as AUTOMAP. \hl{Although Swin Cascade achieved slightly higher SSIM scores, the differences in metrics were minimal, and the sharper delineation of fine structures — particularly around the ventricles — favors the Unrolled reconstruction in visual assessment.}

Additionally, motion artifacts are observed in the fully-sampled reference image, where 44 minutes were required for acquisition. In contrast, motion artifacts are not seen in the reconstructions from accelerated scans, where the acquisitions required $\sim22$ minutes for the 2$\times$ accelerated scan and $\sim11$ minutes for the 4$\times$ accelerated one. This indicates improved subject compliance and reduced probability of motion in shorter scans. Therefore, in addition to demonstrating the reconstruction methods for prospectively undersampled data, this experiment also emphasizes the importance of accelerating low-field MRI scans, which currently suffer from excessive scan durations.

\section{Discussion}

Emerging low-field MRI systems offer portability and low cost but face adoption barriers due to low SNR and long scan times. While CS and DL methods have accelerated high-field MRI \cite{Lustig2007,hammernik2018learning,shimron2023ai}, their use in low-field settings remains limited. Performance can vary widely due to differences in hardware, protocols, and SNR across sites \cite{Arnold2023,webb2023tackling}. Although robustness to data deviations has been studied for high-field MRI \cite{antun2020instabilities,knoll2019assessment}, responses to SNR variation—a hallmark of low-field imaging—have not been systematically explored.  

To address this gap, we conducted a comprehensive evaluation of CS and DL reconstruction frameworks across multiple datasets, acceleration factors, and SNR levels, including raw high-field (3~T) and ultra-low field (6.5~mT) data. We also investigated sampling strategies that exploit the scan-repetition (NEX) dimension to balance acceleration and image quality, and present the first demonstration of the hidden noise problem in low-field MRI.  

\textbf{\textit{Efficient sampling and reconstruction in low-SNR regimes}}. ultra-low field MRI often requires many repetitions to boost SNR via $k$-space averaging, yet this NEX dimension is rarely considered in sampling optimization. We evaluated strategies combining $k$-space undersampling, repetitions, and CS reconstruction within a fixed scan-time budget. This approach consistently outperformed fully sampled IFFT reconstructions, with Poisson Disc undersampling yielding higher quality than Solid Disc masks at the same acceleration, and showed robustness across a wide SNR range—extending earlier high-field fluorine-19 findings \cite{Schoormans2020} to proton imaging in ultra-low field conditions. \hl{Because poor SNR, rather than scan acceleration, is often the primary limitation, image-domain deep learning methods for denoising, artifact removal, and super-resolution offer a valuable complement to $k$-space–based reconstruction \cite{Lau2023pushing,Man2023,Zhao2024,Iglesias2023synthsr}. Although our work focuses on physics-guided reconstruction from undersampled $k$-space, the Unrolled and Swin Cascade models perform substantial image-domain refinement and do not enforce final-stage data consistency, blurring the line between reconstruction and enhancement. Future studies could explore hybrid pipelines that integrate physics-based reconstruction with image-domain super-resolution or contrast harmonization, particularly for low-SNR settings.} \hl{While our experiments targeted 2--4$\times$ acceleration, we note that 4$\times$ undersampling is already challenging for single-coil, ultra-low field MRI; even at high field, DL-based reconstruction often struggles beyond this level \cite{Radmanesh2022}. We therefore focused on acceleration factors that are realistically achievable in current low-field systems, where SNR remains the fundamental constraint.}

\textbf{\textit{Robustness of reconstruction frameworks for low-field MRI}}. We evaluated \hl{four} well-established reconstruction frameworks—Compressed Sensing (CS), data-driven DL (AUTOMAP), physics-guided unrolled networks, and \hl{transformer-based physics-guided architectures (Swin Cascade)}—across different acceleration factors and SNR levels, using both retrospective low-field MRI data synthesized from fastMRI (Figures \ref{fig:fig3_fastmri_metric_curves}, \ref{fig:fig4_fastmri_imgs}) and raw experimental data from our 3~T and 6.5~mT systems (Figures \ref{fig:fig6_retro_3T_ulf}, \ref{fig:fig7_prospective-deployment}).  

\hl{Physics-guided unrolled neural networks outperformed all other methods, including the Swin Cascade, on extremely low-SNR data. Across both retrospective and prospective experiments, the unrolled model produced sharper images, was more robust to noise, and better preserved anatomical detail in ultra-low SNR regimes.} This aligns with high-field MRI studies showing that unrolling iterative optimization algorithms with data consistency blocks can outperform purely data-driven approaches \cite{hammernik2018learning,aggarwal2019modl,hammernik2023physics}. \hl{The Swin Cascade architecture excelled in high-field MRI, consistently achieving the highest NRMSE and SSIM values, likely due to its ability to model long-range spatial dependencies.} \hl{Although Swin Cascade may benefit from further architecture tuning that could lead it to outperform Unrolled reconstructions, it is more computationally demanding. The unrolled network is lightweight, fast to train, and often better suited to low-SNR regimes—highlighting that efficiency can outweigh complexity in real-world low-field settings.} \hl{Overall, convolutional and attention-based physics-guided networks show complementary strengths, with unrolled architectures particularly well-suited to ultra-low field MRI.} CS and AUTOMAP achieved broadly similar performance across datasets and conditions, supporting their continued relevance for low-field reconstruction.

\textbf{\textit{Relation to previous CS and DL studies.}} 
Although CS and DL have been extensively explored for high-field MRI, their application to low-field imaging remains limited. CS has been tested in Overhauser-enhanced systems \cite{Sarracanie2013compressed,Waddington2018}, where polarization transfer boosts SNR—making results less applicable to conventional low-field scanners—and in work targeting distortion correction \cite{Koolstra2021} without assessing noise robustness. \hl{Most prior CS studies have not examined how performance scales across the wide SNR range typical of portable and ultra-low field MRI.} Our study addresses this gap through systematic evaluation of reconstruction stability across SNR regimes and acceleration factors, using both retrospective and prospective experiments.

\hl{Recent DL studies in low-field MRI have focused mainly on post-processing tasks such as super-resolution and denoising \cite{Koonjoo2021,Ayde2022,DeLeeuwdenBouter2022,Islam2023,Iglesias2023,Ayde2024review}, with fewer efforts on direct $k$-space reconstruction.} Notable exceptions include Man \textit{et al.} \cite{Man2023} (partial Fourier sampling) and Zhou \textit{et al.} \cite{zhou2022dual} (spiral acquisitions), neither reflecting the Cartesian undersampling patterns typically used in clinical MRI reconstruction studies. \hl{Other studies, such as Rahman \textit{et al.} \cite{Rahman2025} and Sheng \textit{et al.} \cite{Sheng2024}, introduced Swin Transformer architectures but focused on mid/high-field data without testing robustness to SNR variation or deployment in ultra-low field settings.}

\hl{In contrast, our work uses standard Cartesian acquisitions with variable-density Poisson Disc undersampling to simulate realistic aliasing, and evaluates models on both public data and raw experimental scans at 3~T and 6.5~mT, including prospective deployment.} \hl{This combination of prospective and retrospective validation across field strengths and SNR levels provides a practical benchmark of reconstruction robustness and highlights complementary strengths of convolutional and transformer-based networks.}

\textbf{\textit{The hidden noise problem in low-field MRI}}.  
Recent work has examined instabilities and sensitivities of CS and DL reconstruction methods to specific input-data features \cite{Shimron2022,antun2020instabilities,darestani2021measuring,desai2023noise2recon,wang2024hidden}. Two studies highlighted issues with reference-based image-quality metrics: Shimron \textit{et al.} \cite{Shimron2022} showed that hidden preprocessing in public datasets can bias retrospective evaluations, while Wang \textit{et al.} \cite{wang2024hidden} demonstrated that noise in root sum-of-squares reference images can mis-rank reconstruction algorithms. \hl{This hidden noise problem is distinct from the general limitations of full-reference metrics such as NRMSE or SSIM—it can occur even with a fixed metric if the reference image itself contains noise or artifacts, a common scenario in ultra-low field MRI.}  

Here, we present the first analysis of this issue for low-field MRI. Using our 6.5~mT scanner, we generated two reference images by averaging NEX~=~16 and NEX~=~256 datasets, the latter having higher SNR. Reconstructions from 2$\times$ undersampled data (zero-filling, CS, Unrolled DL, \hl{and Swin Cascade}) were evaluated against each reference. As shown in Figure~\ref{fig:fig5_reference-noise}, high-noise references produced misleading rankings, confirming that algorithm performance can appear better or worse depending solely on reference noise. In agreement with \cite{Shimron2022,wang2024hidden}, this underscores the need for careful reference selection. Potential mitigation strategies include noise augmentation during training \cite{desai2023noise2recon}, noise-aware metrics such as the noncentral chi error \cite{wang2024hidden}, and SNR-focused hardware improvements \cite{webb2023tackling}.

\textbf{\textit{Methods evaluation on real-world prospectively acquired low-field MRI data}}.
The first part of our work (Figures \ref{fig:fig2_fastmri_sampling} to \ref{fig:fig3_fastmri_metric_curves}) was performed using low-field data synthesized from the high-field fastMRI database \cite{Knoll2020}. To validate that our results are not specific to this `fastMRI domain', we deployed our models on raw experimental data acquired in low- and high-field scans (Figures \ref{fig:fig4_fastmri_imgs} to \ref{fig:fig7_prospective-deployment}). Experiments were done using retrospective undersampling and with data acquired with prospective undersampling. The results confirmed our previous findings: \hl{both the unrolled DL and Swin Cascade methods provided high-quality results across all examined field strengths, data types, and acceleration factors.} Notably, for some of the 6.5~mT brain MRI data (Figure \ref{fig:fig6_retro_3T_ulf}), \textit{only} the unrolled DL method was able to remove undersampling artifacts while also maintaining image sharpness; this indicates the powerful capabilities of this approach. \hl{While the Swin Cascade also produced artifact-free reconstructions, it exhibited slightly more noise texture in ultra-low SNR conditions, suggesting that convolutional priors may offer an advantage in extreme low-SNR environments.}

\textbf{\textit{Limitations and future work}}.  
This study focused on single-coil acquisitions, which characterize many low-field MRI systems \cite{Arnold2023,Marques2019}, including our 6.5~mT scanner \cite{Sarracanie2015}. As single-coil setups permit only moderate acceleration, we limited our analysis to 2$\times$–4$\times$ undersampling. Although multi-coil low-field MRI has recently been demonstrated \cite{zhou2022dual}, suitable public datasets are scarce, and extending our work to this setting is left for future research. \hl{Both unrolled and transformer-based frameworks could be adapted to multi-coil data by treating the coil dimension as additional channels without major architectural changes.}  

Public datasets below 0.1~T are especially limited \cite{bell2023sharing}. While the 0.3~T M4Raw database \cite{lyu2023m4raw} provides valuable multi-coil data in the mid-field regime, it differs substantially in SNR characteristics from ultra-low field systems, limiting its direct applicability here. \hl{Our Swin Cascade was trained from scratch, and scaling such models will require larger, standardized datasets matched to the target field strength.}  

Future priorities include developing multi-coil ultra-low field systems, expanding public datasets across field strengths, and exploring advanced training strategies—such as denoising-aware loss functions or SNR-conditioned learning \cite{huang2024nila,lin2024zero}—which may be particularly effective for transformer-based architectures. \hl{Continued refinement of Swin-based and diffusion-driven methods offers a promising route to improving image quality and robustness in portable, low-SNR MRI.}

In summary, this study evaluated classical and deep learning–based reconstruction frameworks for low-field MRI across a range of acceleration factors and SNR levels. \hl{Unrolled convolutional networks and Swin Cascade transformers showed complementary strengths, with unrolled networks outperforming in ultra-low SNR regimes and transformers excelling at high SNR.} We also demonstrated the hidden noise problem in low-field MRI, underscoring the need for careful evaluation practices. These findings can guide the design of acquisition–reconstruction strategies for faster, higher-quality low-field imaging and support emerging applications such as interventional radiology \cite{CampbellWashburn2019} and MRI-guided radiotherapy \cite{Keall2022,Grover2024}.

\vspace{\baselineskip}

\textbf{Acknowledgements:} The authors thank Nicholas Hindley and Helen Ball for providing valuable feedback on a complete draft of this article.

\textbf{Funding:}  D.E.J.W., J.G. and T.B. received support from Australian National Health and Medical Research Council Investigator Grants 2017140 and 1194004. J.G. is supported by an Australian Government Research Training Program scholarship. The information, data, or work presented herein was funded in part by the Advanced Research Projects Agency-Energy (ARPA-E), U.S. Department of Energy, under Award Number DE-AR0000823. The views and opinions of authors expressed herein do not necessarily state or reflect those of the United States Government or any agency thereof. M.S.R. acknowledges the gracious support of the Kiyomi and Ed Baird MGH Research Scholar Award. T.B. acknowledges the support of a Fulbright Future Scholarship, funded by The Kinghorn Foundation, from the Australian–American Fulbright Commission. E.S. is a Horev Fellow and acknowledges the generous support of the Technion's Leaders in Science and Technology program, Zimin Foundation, and Alon Fellowship.

\textbf{Author contributions:} 
Conceptualization: DEJW, ES, and MSR.
Methodology: SSha, NK, SShe, and JG.
Investigation: SShe, TB, JK, AJS, and MSR.
Visualization: DEJW, and ES.
Funding acquisition: DEJW, ES, and MSR.
Project administration: DEJW, and MSR.
Supervision: DEJW, ES, and MSR.
Writing – original draft: DEJW, and ES.
Writing – review \& editing: All.

\textbf{Data availability:} The authors declare that the data supporting the findings of this study are available within the article and its supplementary information. Code necessary to reproduce the results of this paper has been made available at \url{https://github.com/Image-X-Institute/Low_Field_MRI_Recon}. Further raw data are available from the corresponding author upon reasonable request.

\textbf{Competing Interests:} M.S.R. is a founder and equity holder of Hyperfine Inc. MSR is an equity holder of DeepSpin, GmbH. M.S.R. holds two patents relating to the AUTOMAP method used in this work (11,620,772 and 12,272,129).

\bibliographystyle{IEEEtran2025}
\bibliography{ieeebib2024}

\begin{thebibliography}{10}
\providecommand{\url}[1]{#1}
\csname url@samestyle\endcsname
\providecommand{\newblock}{\relax}
\providecommand{\bibinfo}[2]{#2}
\providecommand{\BIBentrySTDinterwordspacing}{\spaceskip=0pt\relax}
\providecommand{\BIBentryALTinterwordstretchfactor}{4}
\providecommand{\BIBentryALTinterwordspacing}{\spaceskip=\fontdimen2\font plus
\BIBentryALTinterwordstretchfactor\fontdimen3\font minus
  \fontdimen4\font\relax}
\providecommand{\BIBforeignlanguage}[2]{{%
\expandafter\ifx\csname l@#1\endcsname\relax
\typeout{** WARNING: IEEEtran.bst: No hyphenation pattern has been}%
\typeout{** loaded for the language `#1'. Using the pattern for}%
\typeout{** the default language instead.}%
\else
\language=\csname l@#1\endcsname
\fi
#2}}
\providecommand{\BIBdecl}{\relax}
\BIBdecl

\bibitem{Sarracanie2015}
M.~Sarracanie, C.~D. Lapierre, N.~Salameh, D.~E.~J. Waddington, T.~Witzel, and
  M.~S. Rosen, ``Low-Cost High-Performance MRI,'' \emph{Sci. Rep.}, vol.~5, pp.
  15\,177--15\,177, 2015.

\bibitem{Kimberly2023}
W.~T. Kimberly \emph{et~al.}, ``Brain imaging with portable low-field MRI,''
  \emph{Nat Rev Bioeng}, vol.~1, pp. 617--630, 2023.

\bibitem{Zhao2024}
Y.~Zhao \emph{et~al.}, ``Whole-body magnetic resonance imaging at 0.05 Tesla,''
  \emph{Science}, vol. 384, no. 6696, 2024.

\bibitem{Salameh2023}
N.~Salameh, D.~J. Lurie, O.~Ipek, C.~Z. Cooley, and A.~E. Campbell-Washburn,
  ``Exploring the foothills: benefits below 1 Tesla?'' \emph{Magnetic Resonance
  Materials in Physics, Biology and Medicine}, vol.~36, no.~3, pp. 329--333,
  2023.

\bibitem{Mazurek2021}
M.~H. Mazurek \emph{et~al.}, ``Portable, bedside, low-field magnetic resonance
  imaging for evaluation of intracerebral hemorrhage,'' \emph{Nature
  Communications}, vol.~12, no.~1, pp. 5119--5119, 2021.

\bibitem{Obungoloch2018}
J.~Obungoloch \emph{et~al.}, ``Design of a sustainable prepolarizing magnetic
  resonance imaging system for infant hydrocephalus,'' \emph{MAGMA}, vol.~31,
  no.~5, pp. 665--676, 2018.

\bibitem{Yuen2022}
M.~M. Yuen \emph{et~al.}, ``Portable, low-field magnetic resonance imaging
  enables highly accessible and dynamic bedside evaluation of ischemic
  stroke,'' \emph{Science Advances}, vol.~8, no.~16, 2022.

\bibitem{Laso2024}
P.~Laso \emph{et~al.}, ``Quantifying white matter hyperintensity and brain
  volumes in heterogeneous clinical and low-field portable MRI,'' in \emph{2024
  IEEE International Symposium on Biomedical Imaging (ISBI)}, 2024, Conference
  Proceedings, pp. 1--5.

\bibitem{sheng2024breast}
S.~Sheng \emph{et~al.}, ``{Breast imaging with an ultra-low field MRI scanner:
  a pilot study},'' \emph{medRxiv}, pp. 2024--04, 2024.

\bibitem{broche2024field}
L.~Broche \emph{et~al.}, ``{Field Cycling Imaging: a novel modality to
  characterise breast cancer at low and ultra-low magnetic fields below 0.2
  T},'' \emph{Communications Medicine}, vol.~4, p. 221, 2024.

\bibitem{webb2023five}
A.~Webb and J.~Obungoloch, ``Five steps to make MRI scanners more affordable to
  the world,'' \emph{Nature}, vol. 615, no. 7952, pp. 391--393, 2023.

\bibitem{Arnold2023}
T.~C. Arnold, C.~W. Freeman, B.~Litt, and J.~M. Stein, ``Low‐field MRI:
  Clinical promise and challenges,'' \emph{Journal of Magnetic Resonance
  Imaging}, vol.~57, no.~1, pp. 25--44, 2022.

\bibitem{Lustig2007}
M.~Lustig, D.~Donoho, and J.~M. Pauly, ``Sparse MRI: The application of
  compressed sensing for rapid MR imaging,'' \emph{Magnetic Resonance in
  Medicine}, vol.~58, no.~6, pp. 1182--95, 2007.

\bibitem{feng2017compressed}
L.~Feng, T.~Benkert, K.~T. Block, D.~K. Sodickson, R.~Otazo, and H.~Chandarana,
  ``Compressed sensing for body MRI,'' \emph{Journal of Magnetic Resonance
  Imaging}, vol.~45, no.~4, pp. 966--987, 2017.

\bibitem{wang2021deep}
S.~Wang, T.~Xiao, Q.~Liu, and H.~Zheng, ``Deep learning for fast MR imaging: A
  review for learning reconstruction from incomplete k-space data,''
  \emph{Biomedical Signal Processing and Control}, vol.~68, p. 102579, 2021.

\bibitem{hammernik2023physics}
K.~Hammernik \emph{et~al.}, ``Physics-Driven Deep Learning for Computational
  Magnetic Resonance Imaging: Combining physics and machine learning for
  improved medical imaging,'' \emph{IEEE Signal Processing Magazine}, vol.~40,
  no.~1, pp. 98--114, 2023.

\bibitem{Heckel2024}
R.~Heckel, M.~Jacob, A.~Chaudhari, O.~Perlman, and E.~Shimron, ``Deep learning
  for accelerated and robust MRI reconstruction,'' \emph{Magnetic Resonance
  Materials in Physics, Biology and Medicine}, vol.~37, no.~3, pp. 335--368,
  2024.

\bibitem{Sarracanie2013compressed}
M.~Sarracanie, B.~D. Armstrong, J.~Stockmann, and M.~S. Rosen, ``High speed 3D
  overhauser-enhanced MRI using combined b-SSFP and compressed sensing,''
  \emph{Magn. Reson. Med.}, vol.~71, pp. 735--745, 2013.

\bibitem{Tamada2014}
D.~Tamada and K.~Kose, ``Two-dimensional compressed sensing using the
  cross-sampling approach for low-field MRI systems,'' \emph{IEEE Transactions
  on Medical Imaging}, vol.~33, no.~9, pp. 1905--1912, 2014.

\bibitem{Koolstra2021}
K.~Koolstra, T.~O’Reilly, P.~Börnert, and A.~Webb, ``Image distortion
  correction for MRI in low field permanent magnet systems with strong B0
  inhomogeneity and gradient field nonlinearities,'' \emph{Magnetic Resonance
  Materials in Physics, Biology and Medicine}, vol.~34, no.~4, pp. 631--642,
  2021.

\bibitem{Hernandez2021}
A.~G. Hernandez \emph{et~al.}, ``Improving image quality in low-field MRI with
  deep learning,'' in \emph{2021 IEEE International Conference on Image
  Processing (ICIP)}, 2021, Conference Proceedings, pp. 260--263.

\bibitem{Le2021deep}
D.~B.~T. Le, M.~Sadinski, A.~Nacev, R.~Narayanan, and D.~Kumar, ``Deep
  Learning–based Method for Denoising and Image Enhancement in Low-Field
  MRI,'' in \emph{2021 IEEE International Conference on Imaging Systems and
  Techniques (IST)}, 2021, Conference Proceedings, pp. 1--6.

\bibitem{lin2024zero}
X.~Lin \emph{et~al.}, ``Zero-Shot Low-Field MRI Enhancement via Denoising
  Diffusion Driven Neural Representation,'' in \emph{International Conference
  on Medical Image Computing and Computer-Assisted Intervention}.\hskip 1em
  plus 0.5em minus 0.4em\relax Springer, 2024, pp. 775--785.

\bibitem{Lau2023pushing}
V.~Lau \emph{et~al.}, ``Pushing the limits of low‐cost ultra‐low‐field
  MRI by dual‐acquisition deep learning 3D superresolution,'' \emph{Magnetic
  Resonance in Medicine}, vol.~90, no.~2, pp. 400--416, 2023.

\bibitem{DeLeeuwdenBouter2022}
M.~L. de~Leeuw~den Bouter, G.~Ippolito, T.~P.~A. O’Reilly, R.~F. Remis, M.~B.
  van Gijzen, and A.~G. Webb, ``Deep learning-based single image
  super-resolution for low-field MR brain images,'' \emph{Scientific Reports},
  vol.~12, no.~1, 2022.

\bibitem{Kim2023d}
S.~Kim, H.~F.~J. Tregidgo, A.~K. Eldaly, M.~Figini, and D.~C. Alexander, ``A 3D
  Conditional Diffusion Model for Image Quality Transfer -- An Application to
  Low-Field MRI,'' p. arXiv:2311.06631, November 01, 2023 2023.

\bibitem{Islam2023}
K.~T. Islam \emph{et~al.}, ``Improving portable low-field MRI image quality
  through image-to-image translation using paired low- and high-field images,''
  \emph{Scientific Reports}, vol.~13, no.~1, p. 21183, 2023.

\bibitem{Lucas2023multi}
A.~Lucas \emph{et~al.}, ``Multi-contrast high-field quality image synthesis for
  portable low-field MRI using generative adversarial networks and paired
  data,'' \emph{medRxiv}, p. 2023.12.28.23300409, 2023.

\bibitem{Ayde2022}
R.~Ayde, T.~Senft, N.~Salameh, and M.~Sarracanie, ``Deep learning for fast
  low-field MRI acquisitions,'' \emph{Scientific Reports}, vol.~12, no.~1, pp.
  1--13, 2022.

\bibitem{Oved2025}
T.~Oved \emph{et~al.}, ``Deep learning of personalized priors from past MRI
  scans enables fast, quality-enhanced point-of-care MRI with low-cost
  systems,'' 2025.

\bibitem{jimeno2022artifactid}
M.~M. Jimeno, K.~S. Ravi, Z.~Jin, D.~Oyekunle, G.~Ogbole, and S.~Geethanath,
  ``ArtifactID: Identifying artifacts in low-field MRI of the brain using deep
  learning,'' \emph{Magnetic resonance imaging}, vol.~89, pp. 42--48, 2022.

\bibitem{Liu2021}
Y.~Liu \emph{et~al.}, ``A low-cost and shielding-free ultra-low-field brain MRI
  scanner,'' \emph{Nature Communications}, vol.~12, no.~1, 2021.

\bibitem{Zhao2024electromagnetic}
Y.~Zhao, L.~Xiao, Y.~Liu, A.~T. Leong, and E.~X. Wu, ``Electromagnetic
  interference elimination via active sensing and deep learning prediction for
  radiofrequency shielding-free MRI,'' \emph{NMR in Biomedicine}, vol.~37,
  no.~7, p. e4956, 2024.

\bibitem{zhou2022dual}
B.~Zhou \emph{et~al.}, ``Dual-domain self-supervised learning for accelerated
  non-Cartesian MRI reconstruction,'' \emph{Medical Image Analysis}, vol.~81,
  p. 102538, 2022.

\bibitem{Man2023}
C.~Man \emph{et~al.}, ``Deep learning enabled fast 3D brain MRI at 0.055
  tesla,'' \emph{Science Advances}, vol.~9, no.~38, p. eadi9327, 2023.

\bibitem{Koonjoo2021}
N.~Koonjoo, B.~Zhu, G.~C. Bagnall, D.~Bhutto, and M.~S. Rosen, ``Boosting the
  signal-to-noise of low-field MRI with deep learning image reconstruction,''
  \emph{Scientific Reports}, vol.~11, pp. 8248--8248, 2021.

\bibitem{Deoni2022}
S.~C.~L. Deoni \emph{et~al.}, ``Development of a mobile low-field MRI
  scanner,'' \emph{Scientific Reports}, vol.~12, no.~1, p. 5690, 2022.

\bibitem{antun2020instabilities}
V.~Antun, F.~Renna, C.~Poon, B.~Adcock, and A.~C. Hansen, ``On instabilities of
  deep learning in image reconstruction and the potential costs of AI,''
  \emph{Proceedings of the National Academy of Sciences}, vol. 117, no.~48, pp.
  30\,088--30\,095, 2020.

\bibitem{darestani2021measuring}
M.~Z. Darestani, A.~S. Chaudhari, and R.~Heckel, ``Measuring robustness in deep
  learning based compressive sensing,'' in \emph{International Conference on
  Machine Learning}.\hskip 1em plus 0.5em minus 0.4em\relax PMLR, 2021, pp.
  2433--2444.

\bibitem{knoll2019assessment}
F.~Knoll, K.~Hammernik, E.~Kobler, T.~Pock, M.~P. Recht, and D.~K. Sodickson,
  ``{Assessment of the generalization of learned image reconstruction and the
  potential for transfer learning},'' \emph{Magnetic resonance in medicine},
  vol.~81, no.~1, pp. 116--128, 2019.

\bibitem{Shimron2022}
E.~Shimron, J.~I. Tamir, K.~Wang, and M.~Lustig, ``Implicit data crimes:
  Machine learning bias arising from misuse of public data,'' \emph{Proceedings
  of the National Academy of Sciences of the United States of America}, vol.
  119, no.~13, p. e2117203119, 2022.

\bibitem{wang2024hidden}
J.~Wang, D.~An, and J.~P. Haldar, ``The “hidden noise” problem in MR image
  reconstruction,'' \emph{Magnetic Resonance in Medicine}, 2024.

\bibitem{Knoll2020}
F.~Knoll \emph{et~al.}, ``fastMRI: A Publicly Available Raw k-Space and DICOM
  Dataset of Knee Images for Accelerated MR Image Reconstruction Using Machine
  Learning,'' \emph{Radiology: Artificial Intelligence}, vol.~2, no.~1, pp.
  e190\,007--e190\,007, 2020.

\bibitem{Zbontar2018}
J.~{Zbontar} \emph{et~al.}, ``{fastMRI: An Open Dataset and Benchmarks for
  Accelerated MRI},'' \emph{arXiv}, p. 1811.08839, 2018.

\bibitem{Uecker2014}
M.~Uecker \emph{et~al.}, ``ESPIRiT - An eigenvalue approach to autocalibrating
  parallel MRI: Where SENSE meets GRAPPA,'' \emph{Magnetic Resonance in
  Medicine}, vol.~71, no.~3, pp. 990--1001, 2014.

\bibitem{Ong2019}
F.~Ong and M.~Lustig, ``SigPy: a python package for high performance iterative
  reconstruction,'' in \emph{27th Annual Meeting of the International Society
  of Magnetic Resonance in Medicine}, 2019, Conference Proceedings, p. 4819.

\bibitem{van2013wu}
D.~C. Van~Essen \emph{et~al.}, ``The WU-Minn human connectome project: an
  overview,'' \emph{Neuroimage}, vol.~80, pp. 62--79, 2013.

\bibitem{Hoult1976}
D.~I. Hoult and R.~E. Richards, ``The Signal-to-Noise Ratio of the Nuclear
  Magnetic Resonance Experiment,'' \emph{J. Magn. Reson.}, vol.~24, pp. 71--85,
  1976.

\bibitem{Koonjoo2017}
N.~Koonjoo, B.~Primavera, J.~P. Stockmann, T.~Witzel, L.~L. Wald, and M.~S.
  Rosen, ``Quadrature Head Coil for Brain Imaging at 6.5 mT,'' in \emph{25th
  Meeting of the International Society of Magnetic Resonance in Medicine},
  2017, Conference Proceedings, p. 2664.

\bibitem{Zhu2018}
B.~Zhu, J.~Z. Liu, B.~R. Rosen, and M.~S. Rosen, ``Image reconstruction by
  domain transform manifold learning,'' \emph{Nature}, vol. 555, no. 7697, pp.
  487--492, 2018.

\bibitem{Zhang2018}
J.~Zhang and B.~Ghanem, ``ISTA-Net: Interpretable Optimization-Inspired Deep
  Network for Image Compressive Sensing,'' \emph{Proceedings of the IEEE
  Computer Society Conference on Computer Vision and Pattern Recognition}, pp.
  1828--1837, 2018.

\bibitem{Shan2023}
S.~Shan \emph{et~al.}, ``Distortion-Corrected Image Reconstruction with Deep
  Learning on an MRI-Linac,'' \emph{Magnetic Resonance in Medicine}, vol.~90,
  pp. 963--977, 2023.

\bibitem{Waddington2023}
D.~E.~J. Waddington \emph{et~al.}, ``Real-time radial reconstruction with
  domain transform manifold learning for MRI-guided radiotherapy,''
  \emph{Medical Physics}, vol.~50, pp. 1962--1974, 2023.

\bibitem{liang2021swinir}
J.~Liang, J.~Cao, G.~Sun, K.~Zhang, L.~Van~Gool, and R.~Timofte, ``SwinIR:
  Image Restoration Using Swin Transformer,'' \emph{arXiv preprint
  arXiv:2108.10257}, 2021.

\bibitem{Rahman2025}
T.~Rahman, A.~Bilgin, and S.~D. Cabrera, ``Multi-channel MRI reconstruction
  using cascaded Swin$\mu$ transformers with overlapped attention,''
  \emph{Physics in Medicine \& Biology}, vol.~70, no.~7, p. 075002, 2025.

\bibitem{Sheng2024}
J.~Sheng \emph{et~al.}, ``Cascade dual-domain swin-conv-unet for MRI
  reconstruction,'' \emph{Biomedical Signal Processing and Control}, vol.~96,
  no. Part A, p. 106623, 2024.

\bibitem{Eo2021}
T.~Eo, H.~Shin, Y.~Jun, T.~Kim, and D.~Hwang, ``Accelerating Cartesian MRI by
  domain-transform manifold learning in phase-encoding direction,''
  \emph{Medical Image Analysis}, vol.~63, 2020.

\bibitem{Sun2019}
L.~Sun, Z.~Fan, X.~Ding, Y.~Huang, and J.~Paisley, ``Region-of-interest
  undersampled MRI reconstruction: A deep convolutional neural network
  approach,'' \emph{Magnetic Resonance Imaging}, vol.~63, no. March, pp.
  185--192, 2019.

\bibitem{kaplan2020scaling}
J.~Kaplan \emph{et~al.}, ``Scaling Laws for Neural Language Models,''
  \emph{arXiv}, p. arXiv:2001.08361, 2020.

\bibitem{liu2021swin}
Z.~Liu \emph{et~al.}, ``Swin Transformer: Hierarchical Vision Transformer using
  Shifted Windows,'' in \emph{2021 IEEE/CVF International Conference on
  Computer Vision (ICCV)}, 2021, pp. 9992--10\,002.

\bibitem{Zhang2018lpips}
R.~Zhang, P.~Isola, A.~A. Efros, E.~Shechtman, and O.~Wang, ``The Unreasonable
  Effectiveness of Deep Features as a Perceptual Metric,'' in \emph{2018
  IEEE/CVF Conference on Computer Vision and Pattern Recognition}, 2018, pp.
  586--595.

\bibitem{Grover2024lpips}
J.~Grover and D.~E.~J. Waddington, ``lpips\_torch2tf,'' \emph{Zenodo}, p.
  https://doi.org/10.5281/zenodo.12747631, 2024.

\bibitem{Wang2004}
Z.~Wang, ``Image Quality Assessment: From Error Visibility to Structural
  Similarity,'' \emph{IEEE Transactions on Image Processing}, vol.~13, pp.
  600--612, 2004.

\bibitem{Waddington2022ismrm}
D.~E.~J. Waddington, E.~Shimron, N.~Hindley, N.~Koonjoo, and M.~S. Rosen,
  ``Accelerating Ultra-Low Field MRI with Compressed Sensing,'' in \emph{31st
  Meeting of the International Society of Magnetic Resonance in Medicine},
  2022, Conference Proceedings, p. 1818.

\bibitem{Knoll2020a}
F.~Knoll \emph{et~al.}, ``Deep-Learning Methods for Parallel Magnetic Resonance
  Imaging Reconstruction: A Survey of the Current Approaches, Trends, and
  Issues,'' \emph{IEEE Signal Processing Magazine}, vol.~37, no.~1, pp.
  128--140, 2020.

\bibitem{Reinke2024}
A.~Reinke \emph{et~al.}, ``Understanding metric-related pitfalls in image
  analysis validation,'' \emph{Nature Methods}, vol.~21, no.~2, pp. 182--194,
  2024.

\bibitem{Waddington2020}
D.~E.~J. Waddington, T.~Boele, R.~Maschmeyer, Z.~Kuncic, and M.~S. Rosen,
  ``High-sensitivity in vivo contrast for ultra-low field magnetic resonance
  imaging using superparamagnetic iron oxide nanoparticles,'' \emph{Science
  Advances}, vol.~6, no.~29, pp. eabb0998--eabb0998, 2020.

\bibitem{hammernik2018learning}
K.~Hammernik \emph{et~al.}, ``{Learning a variational network for
  reconstruction of accelerated MRI data},'' \emph{Magnetic resonance in
  medicine}, vol.~79, no.~6, pp. 3055--3071, 2018.

\bibitem{shimron2023ai}
E.~Shimron and O.~Perlman, ``AI in MRI: Computational frameworks for a faster,
  optimized, and automated imaging workflow,'' p. 492, 2023.

\bibitem{webb2023tackling}
A.~Webb and T.~O’Reilly, ``{Tackling SNR at low-field: a review of hardware
  approaches for point-of-care systems},'' \emph{Magnetic Resonance Materials
  in Physics, Biology and Medicine}, vol.~36, no.~3, pp. 375--393, 2023.

\bibitem{Schoormans2020}
J.~Schoormans, G.~J. Strijkers, A.~C. Hansen, A.~J. Nederveen, and B.~F.
  Coolen, ``Compressed sensing MRI with variable density averaging (CS-VDA)
  outperforms full sampling at low SNR,'' \emph{Physics in Medicine and
  Biology}, vol.~65, no.~4, p. 045004, 2020.

\bibitem{Iglesias2023synthsr}
J.~E. Iglesias \emph{et~al.}, ``SynthSR: A public AI tool to turn heterogeneous
  clinical brain scans into high-resolution T1-weighted images for 3D
  morphometry,'' \emph{Science Advances}, vol.~9, no.~5, p. eadd3607, 2023.

\bibitem{Radmanesh2022}
A.~Radmanesh \emph{et~al.}, ``Exploring the Acceleration Limits of Deep
  Learning Variational Network--Based Two‐Dimensional Brain {MRI},''
  \emph{Radiology: Artificial Intelligence}, vol.~4, no.~6, p. e210313, 2022.

\bibitem{aggarwal2019modl}
H.~K. Aggarwal, M.~P. Mani, and M.~Jacob, ``{{MoDL}: Model Based Deep Learning
  Architecture for Inverse Problems},'' \emph{IEEE Transactions on Medical
  Imaging}, vol.~38, no.~2, pp. 394--405, 2019.

\bibitem{Waddington2018}
D.~E.~J. Waddington, M.~Sarracanie, N.~Salameh, F.~Herisson, C.~Ayata, and
  M.~S. Rosen, ``An Overhauser-enhanced-MRI platform for dynamic free radical
  imaging in vivo,'' \emph{NMR in Biomedicine}, vol.~31, no. May 2018, pp.
  e3896--e3896, 2018.

\bibitem{Iglesias2023}
J.~E. Iglesias \emph{et~al.}, ``Quantitative Brain Morphometry of Portable
  Low-Field-Strength MRI Using Super-Resolution Machine Learning,''
  \emph{Radiology}, vol. 306, no.~3, p. e220522, 2023.

\bibitem{Ayde2024review}
R.~Ayde, M.~Vornehm, Y.~Zhao, F.~Knoll, E.~X. Wu, and M.~Sarracanie, ``MRI at
  low field: A review of software solutions for improving SNR,'' \emph{NMR in
  Biomedicine}, 2024.

\bibitem{desai2023noise2recon}
A.~D. Desai \emph{et~al.}, ``{Noise2Recon: Enabling SNR-robust MRI
  reconstruction with semi-supervised and self-supervised learning},''
  \emph{Magnetic Resonance in Medicine}, vol.~90, no.~5, pp. 2052--2070, 2023.

\bibitem{Marques2019}
J.~P. Marques, F.~F.~J. Simonis, and A.~G. Webb, ``Low-field MRI: An MR physics
  perspective,'' \emph{Journal of Magnetic Resonance Imaging}, vol.~49, pp.
  1528--1542, 2019.

\bibitem{bell2023sharing}
L.~C. Bell and E.~Shimron, ``Sharing data is essential for the future of AI in
  medical imaging,'' \emph{Radiology: Artificial Intelligence}, vol.~6, no.~1,
  p. e230337, 2023.

\bibitem{lyu2023m4raw}
M.~Lyu \emph{et~al.}, ``M4Raw: A multi-contrast, multi-repetition,
  multi-channel MRI k-space dataset for low-field MRI research,''
  \emph{Scientific Data}, vol.~10, no.~1, p. 264, 2023.

\bibitem{huang2024nila}
S.~Huang \emph{et~al.}, ``Noise Level Adaptive Diffusion Model for Robust
  Reconstruction of Accelerated MRI,'' in \emph{Medical Image Computing and
  Computer Assisted Intervention -- MICCAI 2024}, M.~G. Linguraru
  \emph{et~al.}, eds.\hskip 1em plus 0.5em minus 0.4em\relax Cham: Springer
  Nature Switzerland, 2024, pp. 498--508.

\bibitem{CampbellWashburn2019}
A.~E. Campbell-Washburn \emph{et~al.}, ``Opportunities in interventional and
  diagnostic imaging by using high-performance low-field-strength MRI,''
  \emph{Radiology}, vol. 293, no.~2, pp. 384--393, 2019.

\bibitem{Keall2022}
P.~J. Keall \emph{et~al.}, ``Integrated MRI-guided radiotherapy —
  opportunities and challenges,'' \emph{Nature Reviews Clinical Oncology},
  vol.~19, pp. 458--470, 2022.

\bibitem{Grover2024}
J.~Grover \emph{et~al.}, ``Super-resolution neural networks improve the
  spatiotemporal resolution of adaptive MRI-guided radiation therapy,''
  \emph{Communications Medicine}, vol.~4, pp. 64--64, 2024.

\end{thebibliography}

\clearpage

\section*{Supplementary Figures}

\setcounter{figure}{0}
\renewcommand{\figurename}{Supplementary Figure}


\begin{figure}[h]
	\centering
	\includegraphics[width=1\linewidth]{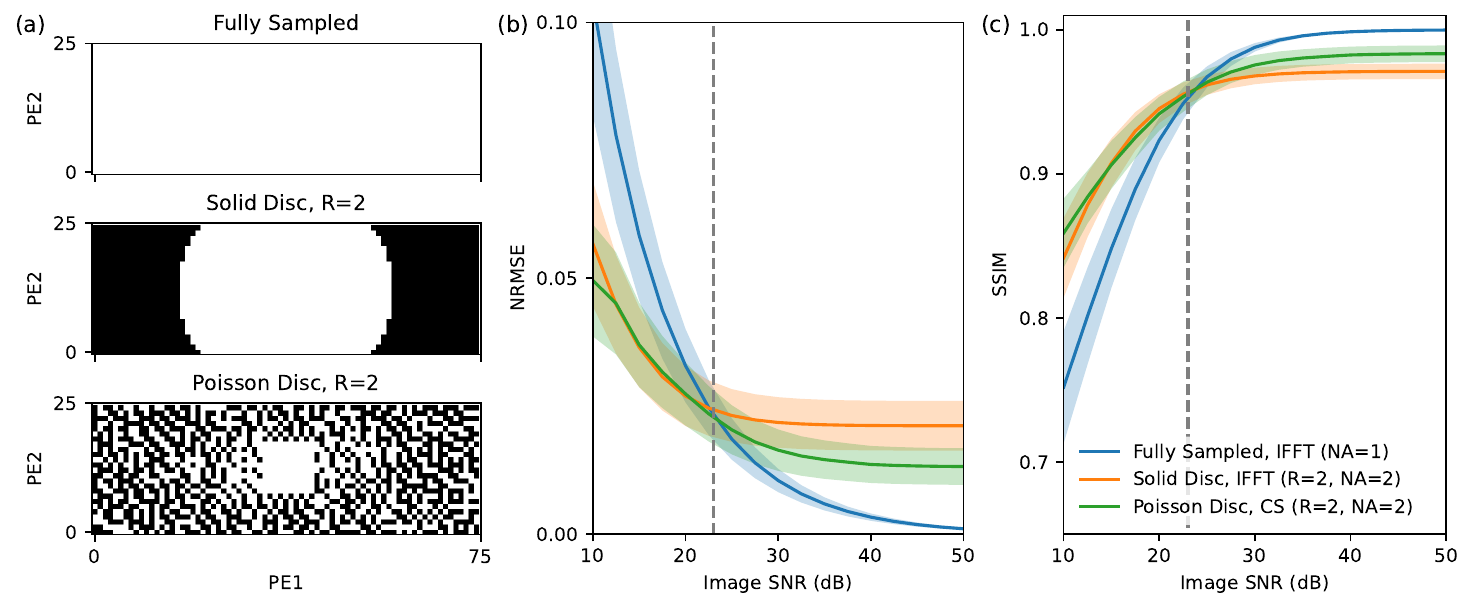}
	\caption{\textbf{Comparing accuracy of image reconstruction methods across  different signal-to-noise ratios, for three sampling schemes.}
		\textbf{(a)} Fully-sampled acquisitions (NEX~=~1) were compared to time-equivalent undersampled approaches ($R=2$, NEX~=~2) using a solid disc pattern and Poisson--disc-like pattern that fills the corners of $k$-space.
		\textbf{(b)} The accuracy of images reconstructed after different sampling approaches was evaluated on fastMRI data using normalized root-mean-square error (NRMSE) and structural similarity metrics (SSIM). Solid lines indicate the mean and shaded regions indicate the standard deviation across 100 brain volumes. The $x$-axis SNR values are calibrated to the Image SNR of the fully-sampled image (i.e. 20~dB is a linear Image SNR of 10, 40~dB is a linear Image SNR of 100).}
	\label{suppfig:fig2_fastmri_sampling}
\end{figure}

\begin{figure}
	\centering
	\includegraphics[width=1\linewidth]{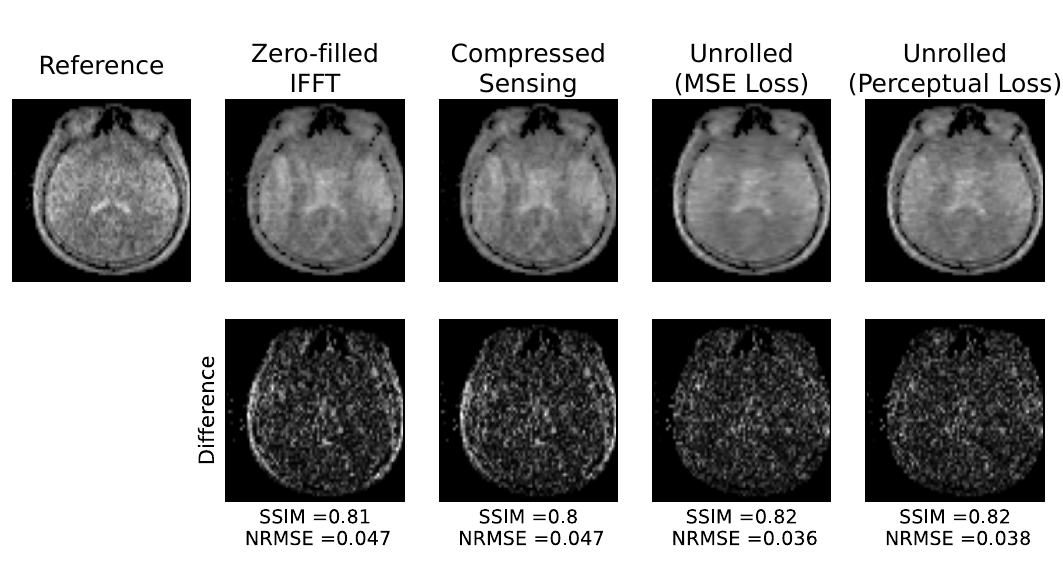}
	\caption{\textbf{Impact of loss functions on reconstruction quality}
		Fully-sampled data were acquired with a number of excitations (NEX) of 16. The NEX=16 data were retrospectively subsampled with an $R=4$ Poisson disc mask and reconstructed with zero-filled Inverse Fast Fourier Transform (IFFT), Compressed Sensing, and unrolled AI techniques. During training, one unrolled AI method utilized Mean Square Error (MSE) loss and the other utilized LPIPS perceptual loss.  Fully-sampled data were reconstructed with IFFT. Structural similarity (SSIM) error and normalized root-mean-square error (NRMSE) metrics were calculated relative to volumes reconstructed with IFFT from both Fully-sampled acquisitions.
		The MSE-trained network reconstructs the brain volume with more smoothing and a lower NRMSE than the LPIPS-trained network. However, the LPIPS-trained network fully removes the aliasing artifacts from the $R=4$ undersampling and leaves other high-frequency features intact, which may potentially be beneficial to clinical interpretation.
	}
	\label{suppfig:perceploss}
\end{figure}

\begin{figure}
	\centering
	\includegraphics[width=1\linewidth]{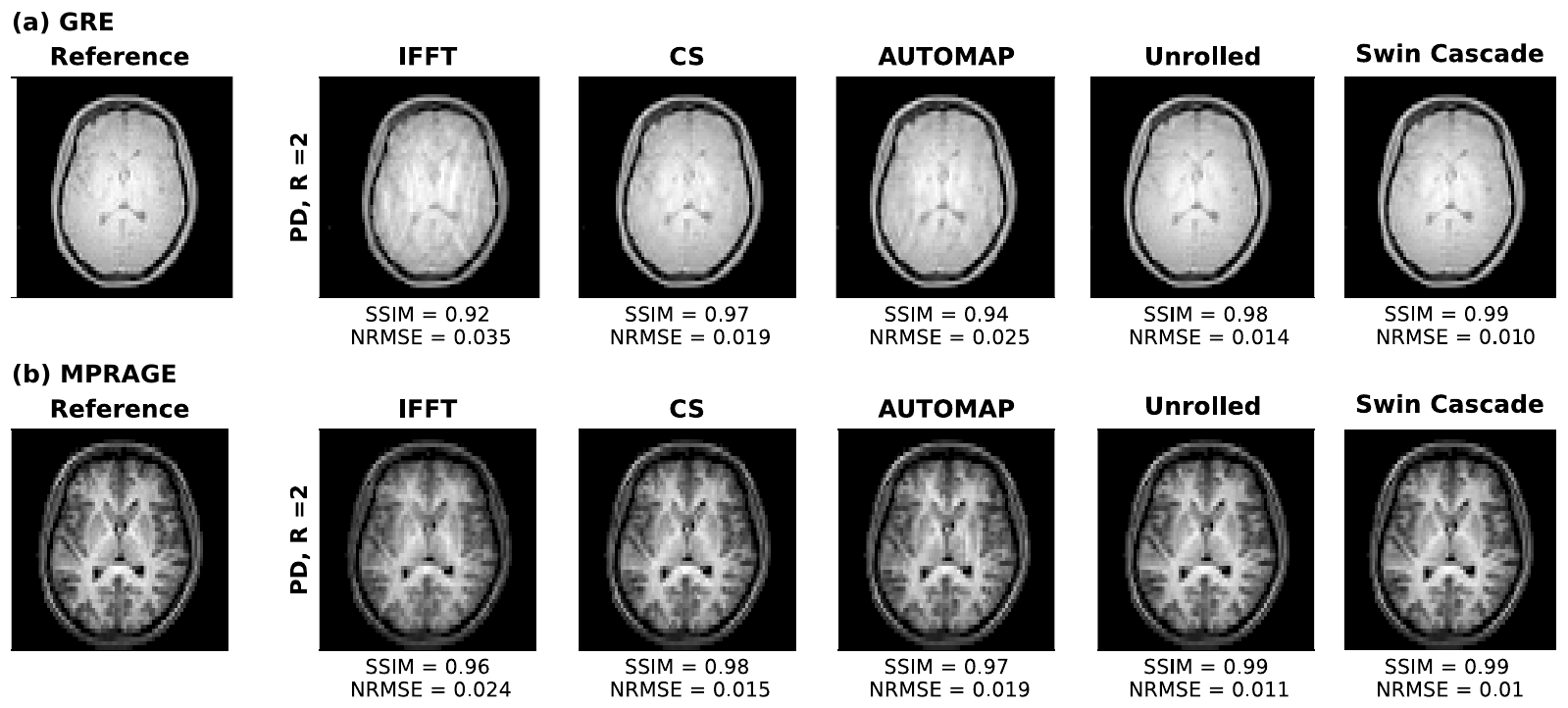}
	\caption{
		\textbf{Further Testing of Reconstruction Methods with High-field Data.}
		Fully-sampled, \textit{single-channel} data were acquired at 3 T using GRE \textbf{(a)} and MPRAGE \textbf{(b)} sequences. A Poisson Disc mask ($R=2$) was retrospectively applied to the $k$-space data. Normalized root-mean-square error (NRMSE) and Structural similarity (SSIM) values are calculated with respect to the reference image. unrolled AI outperforms all other reconstruction techniques. Some aliasing artifacts are visible in the AUTOMAP reconstruction of GRE data, indicating that the data-driven model is not generalizing to this new test data.    
	}
	\label{suppfig:retro_3T_otherseqs}
\end{figure}

\end{document}